\documentclass[twocolumn,tighten]{aastex63}
\pdfoutput=1 %for arXiv submission
\usepackage{amsmath,amstext}
\usepackage{apjfonts} 
\usepackage{afterpage}
\usepackage{xcolor}
\usepackage{threeparttable}
\usepackage{graphicx}
\usepackage{epsfig}

\usepackage{booktabs}
\usepackage{multirow}
\usepackage{float}
\usepackage{mathrsfs}

\setwatermarkfontsize{100px} 

\shortauthors{Satapathy et al.}
\begin{document}

%% LaTeX will automatically break titles if they run longer than
%% one line. However, you may use \\ to force a line break if
%% you desire.

\title{Global Electron Thermodynamics in Radiatively Inefficient Accretion Flows}
\shorttitle{Thermodynamics in Black Hole Accretion}

\author{Kaushik Satapathy}
\affiliation{Department of Physics, University of Arizona, 1118 E. Fourth Street, Tucson, AZ 85721}
\affiliation{School of Physics, Georgia Institute of Technology, 837 State St NW, Atlanta, GA 30332, USA}

\author{Dimitrios Psaltis}
\affiliation{School of Physics, Georgia Institute of Technology, 837 State St NW, Atlanta, GA 30332, USA}

\author{Feryal Ozel}
\affiliation{School of Physics, Georgia Institute of Technology, 837 State St NW, Atlanta, GA 30332, USA}

\begin{abstract}
In the collisionless plasmas of radiatively inefficient accretion flows, heating and acceleration of ions and electrons is not well understood. Recent studies in the gyrokinetic limit revealed the importance of incorporating both the compressive and Alfvenic cascades when calculating the partition of dissipated energy between the plasma species. In this paper, we use a covariant analytic model of the accretion flow to explore the impact of compressive and Alfvenic heating, Coulomb collisions, compressional heating, and radiative cooling on the radial temperature profiles of ions and electrons. We show that, independent of the partition of heat between the plasma species, even a small fraction of turbulent energy dissipated to the electrons makes their temperature scale with a virial profile and the ion-to-electron temperature ratio smaller than in the case of pure Coulomb heating. In contrast, the presence of compressive cascades makes this ratio larger because compressive turbulent energy is channeled primarily into the ions. We calculate the ion-to-electron temperature in the inner accretion flow for a broad range of plasma properties, mass accretion rates, and black hole spins and show that it ranges between $5 \lesssim T_i/T_e \lesssim 40$. We provide a physically motivated expression for this ratio that can be used to calculate observables from simulations of black hole accretion flows for a wide range of conditions. 
\end{abstract}
\section{Introduction}
\label{sec:Intro}

Low-luminosity accretion flows, such as the ones around the black holes at the center of the Milky Way, Sgr~A*, and of the M87 galaxy, belong to a class known as Radiatively Inefficient Accretion Flows \citep{1994ApJ...428L..13N}. They are characterised by low mass-accretion rates, typically $\lesssim 10^{-3}$ times the Eddington accretion rate, and low plasma densities. The collisional timescale between the ions and the electrons in them is significantly larger than the accretion timescale, allowing for the ions and the electrons to co-exist in two different temperatures. Radiative processes are also inefficient, as a result of which the energy dissipated during the accretion process primarily serves to increase the thermal energy carried by the plasma that, in turn, is advected into the black hole~\citep{Narayan1994,Narayan1995a,Narayan1995b}. 

Beyond this general picture of the accretion flow, heating and acceleration of ions and electrons in these collisionless plasmas is not well understood. The turbulence in black-hole accretion flows is thought to be driven by the Magneto-Rotational Instability~\citep[MRI, see ][]{1991ApJ...376..214B}. The MRI creates a turbulent cascade of compressive and Alfven waves. These waves cascade down to small length scales and undergo collisionless dissipation, in the process, channeling a part of their energy into the ions and the rest into the electrons. Even though the dissipation processes operate at multiple length scales, they are only dominant at  scales comparable to the Larmor radius of each species \citep{1998ApJ...500..978Q,1999ApJ...520..248Q,2010MNRAS.409L.104H}. For typical conditions in the accretion flows, these radii are smaller by many orders of magnitude than the macroscopic scales at which MRI is driven. Additional dissipation mechanisms in weakly collisional plasmas might include heating due to velocity-space anisotropies \citep{2018JPlPh..84b7101K,2019MNRAS.486.4013K}, scattering of particles off of velocity-space instabilities \citep{2007ApJ...667..714S,2015ApJ...800...88S}, and magnetic reconnection, which leads to particle acceleration \citep{2016ApJ...826...77B,2018ApJ...853..184B,2017ApJ...850...29R,2019ApJ...873....2R}. 

Analytic studies have shown that, in the inertial range of turbulence, the compressive and Alfvenic cascades decouple from each other at the linear order~\citep{2008PPCF...50l4024S}. These were performed in the gyrokinetic limit, where the gyromotion of particles around the magnetic field were effectively averaged out. The compressive and Alfvenic modes, however, mix at lengthscales comparable to the ion Larmor radius and undergo collisionless damping, which channels a fraction of the turbulent energy into heating the ions. The remaining energy cascades down via a kinetic Alfven wave cascade to smaller lengthscales and dissipates at the electron Larmor radius, heating the electrons. 

Recently, gyrokinetic simulations of collisionless plasmas driven by a combined compressive and Alfvenic turbulent cascade showed that the ion-to-electron heating ratio depends primarily on the plasma $\beta$ parameter and on the ratio $P_{\rm{comp}}/P_{\rm{AW}}$ of the power in the compressive wave cascade to the power in the Alfven wave cascade \citep{2020PhRvX..10d1050K}. This work extends earlier analytic studies of particle heating in purely Alfvenic turbulence \citep{1998ApJ...500..978Q,1999ApJ...520..248Q,2010MNRAS.409L.104H}. 

Because the compressive and Alfvenic wave cascades are decoupled in the inertial range, the partition of turbulent energy between them is determined at the driving scale of the turbulence. As a result, this suggests that the partition of the dissipated energy between ions and electrons at very small scales is determined by the mechanism that drives turbulence at macroscopic scales. In the case of accretion flows around black holes, it is the MRI and the resulting parasitic instability that drive turbulence in the plasma and, therefore, determine the partition of wave energy. The partition of energy under these conditions has not been fully explored. The one set of calculations by \citet{2022JPlPh..88c9011K} were performed using reduced-MHD  simulations of an MRI-driven turbulent shearing flow with a nearly azimuthal background magnetic field and showed that the compressive waves carry at least twice as much power as the Alfven waves. 

In this paper, we utilize these recent developments in our understanding of the physical processes governing the heating of electrons in turbulent flows in order to calculate the ion and electron temperatures in accretion flows and their dependence on plasma parameters. In order to explore a broad range of physical conditions, we employ a covariant semi-analytic model of accretion flows around black holes that we recently developed and calibrated against GRMHD simulations~\citep{2022ApJ...941...88}. We utilize the results of \citet{2020PhRvX..10d1050K, 2022JPlPh..88c9011K} to implement a realistic model of ion and electron heating into this covariant analytic model of accretion, which allows us to calculate the ion-to-electron temperature ratio throughout the accretion flow. 

This approach differs from earlier studies that primarily relied on single-fluid General Relativistic Magneto-HydroDynamic (GRMHD) simulations~\citep[see, e.g.,][]{deVilliers2003,2003ApJ...589..444G,Porth2019} to obtain the bulk properties of the accretion flow and prescribed temperatures in post processing \citep[see, e.g.,][]{2019ApJ...875L...5E,2022ApJ...930L..16A}. In all prescriptions, the temperature ratio $R=T_i/T_e$ is assumed to depend only on the plasma $\beta$ and modeled either as a step function in this parameter~\citep{Chan2015} or as a smooth function between the ratio $R_{\rm{high}}$ in the high-$\beta$ equatorial region and the ratio $R_{\rm{low}}$ in the low-$\beta$ funnel region~\citep{Moscibrodzka2016}. A more first-principles approach is the work of \citet{2015MNRAS.454.1848R}, who derived the energy equations for the electrons and ions individually in GRMHD simulations, using the model for dissipation of Alfvenic turbulence for the partition of heating between the species. Simulations of radiatively inefficient flows using this approach provided additional support to the expectation that the ion-to-electron temperature ratio depends on plasma $\beta$ ~\citep{2017MNRAS.467.3604R,Sadowski2017,2018MNRAS.478.5209C}. 

The paper is organized as follows. In \S \ref{sec:Methods}, we describe our semi-analytical model for the properties of a radiatively inefficient accretion flow around a black hole, elucidating our assumptions related to plasma heating and cooling in constituent subsections. Following this, in \S \ref{sec:Solutions}, we compute solutions for the ion and electron temperatures in the inner accretion flow. We specifically examine the effects of various heating processes, radiative cooling, and our model for the partition of turbulent heat into the ions and the electrons on the ion-to-electron temperature ratio. We summarize our findings in \S \ref{sec:Discussion}. 

\section{Methods}
\label{sec:Methods}
Our goal is to derive a model that describes the spatial distribution of electron and ion temperatures in a radiatively inefficient accretion flow, taking into account particle heating due to the dissipation of Alfvenic and compressive cascades. The key ingredients and assumptions are: 
{\it (i)} a covariant semi-analytic model for the radial dependence of the density, velocity, and internal energy of the accretion flow,  
{\it (ii)} an equation of state for the magnetized plasma consisting of electrons and ions, 
{\it (iii)} a model of the energy dissipation rate in the flow, 
{\it (iv)} macroscopic equations that capture the partition of the dissipated energy between ions and electrons based on the cascade of Alfven and compressive modes, and 
{\it (v)} expressions for the electron cooling rate via synchrotron and Bremsstrahlung processes. 

We consider the dynamics of the accretion flow in a background Kerr metric
\begin{align}
\label{eq:kerr metric}
\begin{split}
ds^2 &= -\left(1-\dfrac{2r}{\Sigma}\right)\ dt^2 -\dfrac{4ar \sin^2{\theta}}{\Sigma}\ dt\ d\phi + \dfrac{\Sigma}{\Delta}\ dr^2 + \Sigma\ d\theta^2  \\ 
&+ \left(r^2 + a^2 + \dfrac{2a^2r \sin^2{\theta}}{\Sigma}\right) \sin^2{\theta}\ d\phi^2, 
\end{split}
\end{align}
\noindent where $\Delta=r^2 - 2r + a^2$, $\Sigma=r^2 + a^2\cos^2{\theta}$, and $a$ denotes the black hole spin. We also use geometrized units with $G=c=M=1$, where G is the gravitational constant, c is the speed of light, and $M$ is the mass of the black hole.  

We describe each element of the model in the following subsections. 

\subsection{A Covariant Model for the Accretion Flow}

The global dynamics of the ions and electrons in an accretion flow where ion-electron collisions are negligible can be described by conservation of mass and conservation of energy independently in each species \citep[see, e.g.,][]{2015MNRAS.454.1848R}. The mass conservation is given by 

\begin{equation}
\label{eq:mass conservation}
\nabla_\mu \left(\rho_{(s)}u_{(s)}^{\mu}\right) = 0,
\end{equation}

\noindent and the conservation of energy can we written as

\begin{equation}
\label{eq:energy conservation}
u_{\nu(s)} \nabla_\mu \left(T_{(s)}^{\mu\nu}\right) = 0.
\end{equation}

\noindent In equations~\eqref{eq:mass conservation} and \eqref{eq:energy conservation}, $\rho$ is the density, $u^\mu$ is the four velocity, $T^{\mu\nu}$ is the stress energy tensor, and the label $s$ in parenthesis for each quantity indicates the species, $i$ (ion) or $e$ (electron). 

We consider a charge-neutral hydrogen plasma such that number densities of electrons and ions are equal, $n_e \approx n_i$, and are related to the mass densities by $\rho_e = m_e n_e$ and $\rho_i = m_i n_i$, where $m_e$ and $m_i$ are the electron and ion masses, respectively. We further assume that both the ions and electrons follow the same flow four-velocity, i.e., $u^\mu_i \approx u^\mu_e \equiv u^\mu$ and that all flow quantities are azimuthally symmetric.

The stress energy tensor in the plasma can be written as 
\begin{align}
\label{eq:stress energy tensor}
\begin{split}
&T^{\mu\nu} = \left(P_i + P_e + \dfrac{B^2}{8\pi}\right) g^{\mu\nu} + \\ &\left(\rho_i + \rho_e + \epsilon_i + \epsilon_e + P_i + P_e + \dfrac{B^2}{4\pi} \right)u^\mu u^\nu + b^{\mu}b^{\nu}+ t^{\mu\nu}.
\end{split}
\end{align}
\noindent Here, $B$ is the magnetic field strength, $\epsilon_{(s)}$ is the internal energy, $\rho_{(s)}$ is the density, and $b^{\mu}$ represents the magnetic field four vector \citep[see][]{2015MNRAS.454.1848R}. The viscous stresses acting upon the plasma are accounted for in the tensor $t^{\mu\nu}$. We define $P_{(s)}$ as the effective pressure of each species to include the effects of the turbulent fluctuations, in addition to the thermal motions of the particles. We address our definition of pressure in greater detail in sub-section \S \ref{subsection: equation of state}.

We further reduce the dimensionality of our model by averaging azimuthally over $2\pi$ and vertically over the scale height of the disk in all fluid quantities so that they are left with only a radial dependence in steady state \citep[see][]{2022ApJ...941...88}. With this integration, the equation for mass conservation becomes 
\begin{equation}
\label{eq:mass accretion rate}
4\pi\left(\dfrac{h}{r}\right)\sqrt{-g}\ \rho(r)\ u^r(r) = -\dot{M},  
\end{equation}
\noindent where $\dot{M}$ is the mass accretion rate in the disk and $h/r$ represents its scale height.  We refer to all variables defined in this way as height-averaged fluid quantities. 

Hereafter, we will express the mass accretion rate in terms of the Eddington critical rate
\begin{equation}
M_{\rm E}= \frac{4 \pi G M m_p} {\eta c \sigma_T},
\end{equation}
where $\sigma_T$ is the Thomson cross section and we set the radiative efficiency of the accretion flow $\eta$ to be 0.1.
We also set the scale height by imposing local vertical pressure balance at each radius~\citep{2022ApJ...941...88}, i.e.,
\begin{equation}
\frac{h}{r} = \frac{1}{r u^\phi} \left(\frac{P_i+P_e}{\rho_i+\rho_e}\right)^{1/2}\;,
\label{eq:hr}
\end{equation}
where $u^\phi$ is the azimuthal angular velocity of the flow, assumed to be equal to the local Keplerian value.

Using the form of the stress energy tensor defined in Equation \eqref{eq:stress energy tensor}, we simplify the conservation of energy in Equation \eqref{eq:energy conservation} to obtain

\begin{equation} 
\label{eq:radial ion energy conservation}
u^r \dfrac{d\epsilon^{\rm{eff}}_i}{dr} - u^r \dfrac{\epsilon^{\rm{eff}}_i + P_i}{\rho_i} \dfrac{d\rho_i}{dr} = Q_i  - Q_{ie}
\end{equation}
\noindent for the ions, and 
\begin{equation} 
\label{eq:radial electron energy conservation}
u_e^r \dfrac{d\epsilon_e^{\rm eff}}{dr} - u_e^r \dfrac{\epsilon_e^{\rm eff} + P_e}{\rho_e} \dfrac{d\rho_e}{dr} = Q_e + Q_{ie} - \Lambda_e 
\end{equation}
\noindent for the electrons. Here, $Q_i$ is the amount of the energy dissipated per unit volume in the plasma that goes into heating the ions, $Q_e$ is the amount of heat that is channeled into the electrons, and $Q_{ie}$ is the energy exchange rate between the ions and the electrons. Because electrons in the plasma cool through synchrotron and Bremsstrahlung radiation, we incorporate the appropriate cooling rate $\Lambda_e$ for the electrons. 

Because of the azimuthal and vertical averaging of the equations, as well as the implicit averaging over the coherence time of the MHD turbulence, the various components of the fluid velocity describe only the overall motions of the plasma and not its fluctuating turbulent motions. Due to the same averaging, both the pressures, $P_i$ and $P_e$, and the effective internal energies, $\epsilon^{\rm{eff}}_i$ and $\epsilon^{\rm{eff}}_e$, include contributions from the fluctuating turbulent plasma motions. Note that the magnetic field does not enter explicitly the energy equations.

Following the approach in \citet{2022ApJ...941...88}, we write the radial velocity outside of the innermost stable circular orbit (ISCO) as \begin{equation}
\label{eq:radial plasma velocity}
u^r = -\eta \left(\dfrac{r}{r_{\rm{ISCO}}}\right)^{-n_R}, 
\end{equation}
where $\eta$ and $n_R$ are parameters that aim to capture the efficiency of the mechanism that drives angular momentum transport in the flow. By exploring a range for these parameters, we allow our model to be general enough to encompass different plausible dissipation profiles.

Using this model for the radial velocity, we can easily apply mass conservation (equation~[\ref{eq:mass accretion rate}]) and obtain a radial profile for density, which yields another power law function in radial distance. 

\subsection{The equation of state of the magnetized ion-electron plasma}
\label{subsection: equation of state}
In order to infer the temperatures of each species, we employ an ideal gas equation of state for the ion and electron thermal pressures. Adding the contributions of the fluctuating turbulent motions, to which we assign an r.m.s.\ velocity of $u_{\rm turb}$, we write the effective pressure of each species as
\begin{equation}
\label{eq:equation of state}
P_{(s)} = n \left(k_B T_{(s)}+ m_{(s)} u_{\rm turb}^2\right)\;. 
\end{equation}
Under the reasonable assumption that the r.m.s.\ turbulent velocities scale with the local Alfv\'en velocity, $u_{\rm turb}^2= \zeta u_{\rm A}^2=\zeta B^2/(4\pi m_{\rm i}n)$, where $\zeta$ is a factor of order unity, the equation for the effective pressure of the electrons becomes
\begin{eqnarray}
\label{eq:equation of state electrons}
P_{e} &=& n k_B T_{e}+ \zeta \left(\frac{m_e}{m_i}\right) \frac{B^2}{4\pi n k_B T_e} n k_B T_e\nonumber\\
&\simeq & n k_B T_{e} \left[1+ \zeta \left(\frac{m_e}{m_i}\right)
\frac{2 R}{\beta}\right]\nonumber\\
&\simeq& n k_B T_{e}\;.
\end{eqnarray}
In contrast, the equation for the effective pressure of the ions becomes
\begin{eqnarray}
\label{eq:equation of state ions}
P_i &=& n k_B T_i+ \zeta \frac{B^2}{4\pi n k_B T_i}n k_B T_i=n k_B T_{i}\left(1+\frac{2 \zeta}{\beta}\right)\;. 
\end{eqnarray}

Similarly, the effective internal energy becomes 
\begin{equation}
\label{eq: internal energy electrons}
\epsilon^{\rm eff}_e \simeq 
\dfrac{n k_B T_e}{\gamma_e - 1}
\end{equation}  
for the electrons and 
\begin{eqnarray}
\label{eq: internal energy ions}
\epsilon^{\rm{eff}}_i &= &\frac{n k_B T_i}{\gamma_i - 1}\left(1 + \dfrac{2\zeta}{\beta} \right) \\
& \equiv&  \dfrac{n k_B T_i}{\gamma_i^{\rm{eff}}-1}
\end{eqnarray} 
for the ions. 
In all these equations, we have defined the plasma parameter as $\beta = 8\pi P_i/B^2$. 

Allowing for trans-relativistic behavior of the ions and electrons, we express the temperature dependence of the adiabatic index of a gas as \citep[see][]{2011irsm.book.....H}
\begin{equation}
\label{eq:gamma trans relativistic}
\gamma\left(\Theta\right) = \dfrac{H-1}{H-1-\Theta},
\end{equation}
where $H = K_3(1/\Theta)/K_2(1/\Theta)$. In this equation, $K_n$ denotes the modified Bessel function of the second kind at $n^{\rm{th}}$ order and the dimensionless temperature for the two species is given by $\Theta_{(s)} \equiv k_B T_{(s)}/m_{(s)}c^2$. 

\subsection{Plasma Heating}

The viscous stresses present in the accretion flow channel the gravitational potential energy into turbulent wave energy that resides as MHD waves in the disk. These MHD waves, in principle, undergo dissipation at microscopic length-scales and channel their energy into the thermal energies of the plasma species (ions and electrons). Hence, the heating rates of the individual species, $Q_i$ and $Q_e$, are determined by two factors:
\noindent \textit{(i)} the total energy dissipated in the plasma due to viscous stresses (defined as $Q \equiv Q_i + Q_e$) and 
\noindent \textit{(ii)} the partition of the total dissipated energy into the ions and electrons (defined as $f \equiv Q_i/Q_e$).

In addition to direct heating, there exists a channel of energy exchange between the ions and the electrons through Coulomb collisions ($Q_{ie}$, see equations [\ref{eq:radial ion energy conservation}] and [\ref{eq:radial electron energy conservation}]), which is given by \citep[see ][]{1984ApJ...280..319C}

\begin{equation}
\label{eq:coulomb coupling}
Q_{ie} = 3.83 \times 10^{-19} n_e^2 \left(\Theta_i - \frac{m_e}{m_i}\Theta_e\right) \left(\Theta_e + \Theta_i\right)^{-3/2}.
\end{equation}

\subsubsection{Energy Dissipation}
\label{sec:viscous heating}

Following~\citet{1998ApJ...498..313G} and \citet{2022ApJ...941...88}, we write the total rate of dissipation in the plasma arising from the viscous stresses $t^{\mu\nu}$ as 
\begin{equation}
\label{eq:heating stress shear}
Q = -t^{\mu\nu}\sigma_{\mu\nu}\, ,
\end{equation}
\noindent where $\sigma_{\mu\nu}$ is the covariant shear tensor. This calculation is simplified in the local rest frame where only the $r\phi$ component of the stress is non-negligible. 

In the non-relativistic limit, the $r\phi$ component of the strain, $\sigma_{(r)(\phi)}$, becomes \citep{2022ApJ...941...88}
\begin{equation}
\label{eq:sigma_r_phi}
\sigma_{(r)(\phi)} \approx \frac{\sqrt{g_{\phi\phi}}}{2}\times \dfrac{d\Omega}{dr}, 
\end{equation}
where $\Omega$ is the angular velocity given by 
\begin{equation}
\label{eq:omega definition}
\Omega = \sqrt{-\dfrac{g_{tt,r}}{g_{\phi\phi,r}}}
\end{equation}
The same component for the viscous stress tensor, $t_{(r)(\phi)}$, is given by 
\begin{equation}
\label{t_r_phi}
t_{(r)(\phi)} \approx \sqrt{\dfrac{g_{rr}}{g_{\phi\phi}}} t^r_\phi. 
\end{equation}
\noindent The mixed $r\phi$ component of the viscous stress tensor, $t^r_\phi$, can be obtained from the conservation of angular momentum
\begin{equation}
\label{eq:angular momentum conservation}
4\pi\left(\dfrac{h}{r}\right)r^2 t^r_{\phi} = \dot{M}\left(L_z-j\right), 
\end{equation}
\noindent where $L_z$ is the angular momentum of the accreting material and $j$ is an eigenvalue. Following \citet{2022ApJ...941...88}, we parametrize $j$ as a fraction of the angular momentum at the ISCO as 
\begin{equation}
\label{eq:eigenvalue j parametrization}
j = \lambda L_z(r_{\rm{ISCO}}). 
\end{equation}

Combining Equations \eqref{eq:sigma_r_phi}-\eqref{eq:eigenvalue j parametrization}, we write the plasma heating rate $Q$ as 
\begin{equation}
\label{eq:expression for Q}
Q = \dfrac{\dot{M}}{4\pi r^2(h/r)}\sqrt{g_{rr}}\dfrac{d\Omega}{dr}\left\{L_z-\lambda L_z(r_{\rm{ISCO}})\right\}.
\end{equation}
\noindent We further simplify this expression by approximating $L_z\approx g_{\phi\phi}\Omega$, valid to leading order in the Schwarzschild metric.

\subsubsection{Partition of Heat}
\label{sec:partition of heating}

The wave energy in the turbulent plasma is driven at macroscopic scales by the magnetorotational instability and cascades down to much smaller scales that are comparable to ion and electron Larmor radii, where it is channeled into thermal energy of the particles. Gyrokinetic approaches, in which it is possible to average out the fast gyro-motions of particles around a mean magnetic field, have shown that the wave energy resides in compressive and Alfven wave cascades that are decoupled in the inertial range \citep[see discussion in \S \ref{sec:Intro}, ][]{2008PPCF...50l4024S}. The length scales where the wave dissipation via wave-particle resonances (such as Landau damping) dominate correspond to the Larmor radii of the particles. Hence, at the length scale of the ion Larmor radius, a fraction of the wave energy is dissipated into the ions, while the remainder cascades down to the smaller electron Larmor radius, effectively partitioning the heat between ions and electrons. 

\begin{figure}
    \includegraphics[width=0.99\linewidth]{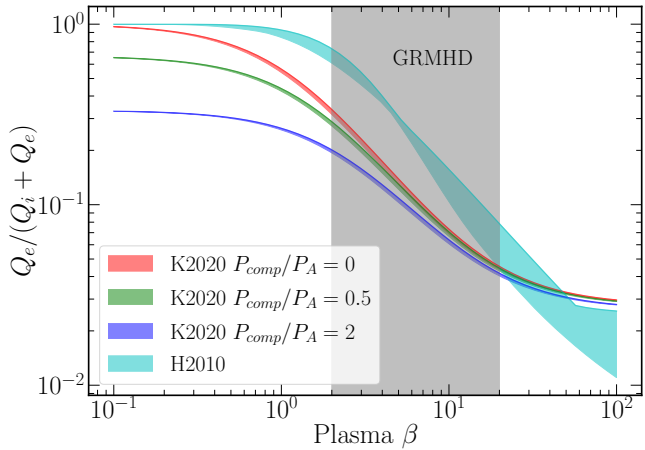}
    \caption{The electron heating fraction, $Q_e/(Q_i + Q_e)$, as a function of the plasma $\beta$, as calculated by \citet{2020PhRvX..10d1050K} for different values of the relative driving power of compressive and Alfvenic waves, 
    $P_{\rm comp}/P_{\rm A}$. The width of each curve represents the range of heating ratios for different ion-to-electron temperature ratios ($1\le T_i/T_e\le 100$) and illustrates the weak dependence on this quantity. The cyan curve corresponds to the earlier analytical calculation by \citet{2011ApJ...738...40H}, which did not incorporate compressive waves. The typical values of plasma $\beta$ found in the midplane regions of GRMHD simulations is shown by the shaded region in gray.}
    \label{fig:howes and kawazura model}
\end{figure}

In this paper, we utilize the results of \citet[][hereafter K2020]{2020PhRvX..10d1050K}, who computed numerically the ion-to-electron heating ratio in a turbulent cascade by means of gyrokinetic simulations, accounting for mode-mixing in the presence of both compressive and Alfvenic cascades at the driving scales of the turbulence \citep[see][for a rigorous description of gyrokinetic turbulent cascades]{2006ApJ...651..590H,2008PPCF...50l4024S}. Allowing for an arbitrary relative power $P_{\rm{comp}}/P_{\rm{AW}}$ between the compressive and Alfvenic cascades and exploring different background plasma conditions, they obtained an approximate empirical expression for the ion-to-electron heating ratio given by
\begin{equation}
\label{eq:kawazura model}
%f\left(\beta,\dfrac{T_i}{T_e}, \dfrac{P_{\rm{comp}}}{P_{\rm{AW}}}\right) \equiv 
\frac{Q_i}{Q_e} = \dfrac{35}{1 + (\beta/15)^{-1.4}\exp\{{-0.1/(T_i/T_e)}\}} + \dfrac{P_{\rm{comp}}}{P_{\rm{AW}}}. 
\end{equation}
Figure~\ref{fig:howes and kawazura model} shows the dependence of the electron heating fraction $Q_e/(Q_e+Q_i)$ on plasma $\beta$, allowing for different values of $P_{\rm{comp}}/P_{\rm{AW}}$. The dependence of the heating fraction on the ion-to-electron temperature ratio is captured by the width of each curve, for values in the range $1\le T_i/T_e\le 100$, and is very weak.

As discussed earlier, the partition of heat by means of wave-particle interactions relies on the separation between the ion and electron Larmor radii. The latter is set by the ratio of temperatures of the species. The temperature ratios in radiatively inefficient flows are such that the ion Larmor radius is always much larger than the electron Larmor radius. As a result, the temperature ratio does not have a strong influence on the partition of heating and $Q_i/Q_e$ is primarily determined by the plasma $\beta$. 

In the limit of high plasma $\beta$, irrespective of the nature of the turbulent cascade, strong ion Landau damping causes the majority of the heat to be channeled into thermal energy of the ions. On the other hand, in the limit of low plasma $\beta$, i.e., when the plasma is magnetically dominated, the Alfven-wave speeds are relatively high. Therefore, the population of ions available to resonate with the Alfven waves significantly drops and almost all of the Alfven wave energy goes into heating the electrons instead. The ions, in turn, only heat up from the turbulent energy that resides in the compressible modes \citep[][]{2019JPlPh..85c9003S}. This is the reason why, at the low-$\beta$ limit, the ion-to-electron heating ratio becomes equal to $P_{\rm{comp}}/P_{\rm{AW}}$. 

For comparison, we also consider the earlier results for the ion-to-electron heating ratio by \citet[][hereafter H2010]{2010MNRAS.409L.104H} who calculated analytically the linear dissipation of a purely Alfvenic cascade. The ion-to-electron heating ratio described by this model is given by 
\begin{equation}
\label{eq:howes model}
\frac{Q_i}{Q_e} = c_1 \frac{c_2 + \beta^p}{c_3 + \beta^p}\sqrt{\frac{m_i T_i}{m_e T_e}} \exp{(-1/\beta)}\;, 
\end{equation}
where $c_1 = 0.92$, $c_2 = 1.6/(T_i/T_e)$, $c_3 = 18 + 5 \log{(T_i/T_e)}$, and $p = 2 - 0.2 \log{(T_i/T_e)}$, and is also shown in Figure~\ref{fig:howes and kawazura model}. Because of the absence of the compressive modes, the low-$\beta$ limits of these calculation channels all the dissipated energy to the electrons. Moreover, the linear approximations inherent to that calculation result in a somewhat stronger dependence of the heat partition on the ion-to-electron temperature ratio. In the next section, we will discuss how these assumptions impact the resulting ion-to-electron temperature ratio.

The above mentioned strong dependence of the heating fraction on the composition of turbulent cascade ($P_{\rm{comp}}/P_{\rm{AW}}$) is quite significant for the plasmas found in radiatively inefficient disks. GRMHD simulations carried out in a standard and normal evolution (SANE) configuration estimate a midplane plasma $\beta \sim 10$ in the inner accretion flow and $\beta \lesssim 1$ in the magnetically dominated funnel~\citep[see ][]{Porth2019}. The same quantity for a simulation of a magnetically arrested disk (MAD) lies in the range $2 \leq \beta \leq 10$, even with the equatorial accretion flow~\citep{Sadowski2015}. As shown in Figure \ref{fig:howes and kawazura model}, in this range of values for plasma $\beta$, the electron heating fraction shows an appreciable dependence on the ratio of compressive to Alfven turbulent energy $P_{\rm{comp}}/P_{\rm{AW}}$. 

The ratio of the compressible to Alfvenic wave energy in MRI-driven turbulence is not well understood. To date, there has been one set of shearing box simulations of MRI-driven turbulence at the reduced MHD limit in a nearly azimuthal magnetic field configuration, for which the energy injection rates in the compressive and Alfvenic cascades have been computed \citep{2022JPlPh..88c9011K}. This calculation shows that, in the regime of $0.1 \leq \beta \leq 10$, compressive waves carry twice as much energy as Alfven waves, almost independent of plasma $\beta$ ($P_{\rm{comp}}/P_{\rm{AW}}=2$). We will use this estimate of the composition of the cascade at driving scales as a fiducial model but also explore the impact of this value on the thermodynamics of the flow by allowing a variety of other $P_{\rm{comp}}/P_{\rm{AW}}$ values.

\subsection{Electron Cooling}
\label{sec: electron cooling subsection}

In a radiatively inefficient flow, the electrons cool primarily via synchrotron and Bremsstrahlung radiation \citep[see][]{2014ARA&A..52..529Y}. The synchrotron emissivity ($j_\nu$) for temperatures above the trans-relativistic regime can be written as \citep[][]{1996ApJ...465..327M}

\begin{equation}
\label{eq:synchrotron cooling}
j_\nu = \dfrac{n_e e^2}{\sqrt{3}c K_2(1/\Theta_e)}\nu M(x_M), 
\end{equation}
\noindent where $K_2$ is the modified Bessel function of the second kind and second order, $x_M = 2\nu/(3\nu_b \Theta^2)$, and $\nu_b = eB/(2\pi m_e c)$ is the cyclotron frequency. The functional form of $M(x_M)$ is provided in \citet{1996ApJ...465..327M}. 

When the optical depth is low, the electron cooling rate, $\Lambda_e$, is simply the emissivity integrated over all frequencies and is given by

\begin{equation}
\label{eq:electron cooling rate}
\begin{split}
\Lambda_e &= \int_0^\infty j_\nu d\nu = \dfrac{9\sqrt{3}c\sigma_T}{128\pi^3}\left(n_e B^2\right)\left[\dfrac{\Theta_e^4}{K_2(1/\Theta_e)}\right]\\
&\qquad\qquad \times\int_0^\infty dx_M M(x_M) x_M,
\end{split}
\end{equation}

\noindent where $\sigma_T$ is the Thomson cross section. We obtain the magnetic field strength in Equation~\eqref{eq:electron cooling rate} from the plasma $\beta$ as $B^2 = 8\pi n k_B T_e/\beta$. The integral in $x_M$ is computed numerically as

\begin{equation}
\label{eq:definition of C}
C \equiv \int_0^\infty dx_M x_M M(x_M) \approx 20.\;.
\end{equation}

The electron cooling rate via Bremsstrahlung processes has two contributions. For ion-electron interactions, it is given by
\begin{equation}
\label{eq:bremmstrahlung ion-electron}
\Lambda_{B}^{ie} = 1.48 \times 10^{-22}\ n_e^2 \ F(\Theta_e), 
\end{equation}
where
\begin{equation}
\label{eq:bremmstrahlung ion-electron F}
F(\Theta_e) = \left\{
        \begin{array}{ll}
            4\left(\dfrac{2\Theta_e}{\pi^3}\right)^{1/2} \times  \left(1 + 1.781\ \Theta_e^{1.34}\right)& \quad \Theta_e < 1 \\
            \dfrac{9\Theta_e}{2\pi} \left\{\ln(1.123\ \Theta_e + 0.48) + 1.5 \right\}& \quad \Theta_e > 1
        \end{array}
    \right., 
\end{equation}
\noindent and for electron-electron interactions, it is written as
\begin{equation}
\label{eq:bremmstrahlung electron-electron}
\Lambda_{\rm{B}}^{\rm{ee}}(\Theta_e) = \left\{
        \begin{array}{ll}
            2.56\times 10^{-22}\ n_e^2 \times\ \Theta_e^{3/2}\\ \times \left(1+1.1\Theta_e+\Theta_e^2-1.25\Theta_e^{5/2}\right) & \quad \Theta_e < 1 \\
            3.40\times 10^{-22}\ n_e^2  \\ \times\ \Theta_e\left\{\ln\left(1.123\Theta_e + 1.28\right)\right\}& \quad \Theta_e > 1.
        \end{array}
    \right.
\end{equation}

We note that, while the rate of viscous heating is directly proportional to the plasma density, the resultant energy transfer rates due to Coulomb collisions, Bremsstrahlung cooling, and synchrotron cooling are proportional to the square of the plasma density, introducing an overall scale which is linear in plasma density into the equations \eqref{eq:radial ion energy conservation} and \eqref{eq:radial electron energy conservation}. This is why the overall mass accretion rate does not drop out of Equations~\eqref{eq:radial ion energy conservation} and \eqref{eq:radial electron energy conservation} and needs to be specified, unlike the case in adiabatic GRMHD simulations. 

\section{Solutions for Ion and Electron Temperatures}
\label{sec:Solutions}

In this section, we solve numerically the energy equations for ions and electrons (Eqs. \ref{eq:radial ion energy conservation} and \ref{eq:radial electron energy conservation}), to obtain the radial dependence of their temperatures. We adopt the fiducial setup for the ion-to-electron heating ratio with $P_{\rm{comp}}/P_{\rm{AW}} = 2$. In order to solve the system of differential equations, we initialize the ions and electron temperatures at a radial distance of $\sim 2\times 10^5 GM_{\rm{BH}}/c^2$ , assigning a sub-virial non-relativistic temperature to both species. In the following subsections, we examine the impact on the solutions for the ion and electron temperatures of the efficiency of turbulent and Coulomb heating, of radiative cooling, as well as of the model parameters for the partition of energy dissipation. We discuss the contribution of General Relativistic effects in the Appendix.

\subsection{Effect of Turbulent Heating}
\label{sec:compressional heating subsection}

We first solve for the ion and electron temperatures in a flow with a mass accretion rate $\dot{M} \ll 10^{-7} \dot{M}_{E}$. In this limit of extremely low mass accretion rate, turbulent heating dominates over the effects of Coulomb heating, while radiative cooling is negligible. We also set the strength of the magnetic field in the plasma by choosing the fiducial value $\beta=5$, the radial velocity profile by setting its power-law index to $n_R = 1.5$, and the value of $\zeta = 0.2$ for all our calculations. 

\begin{figure}[t]
    \includegraphics[width=0.99\linewidth]{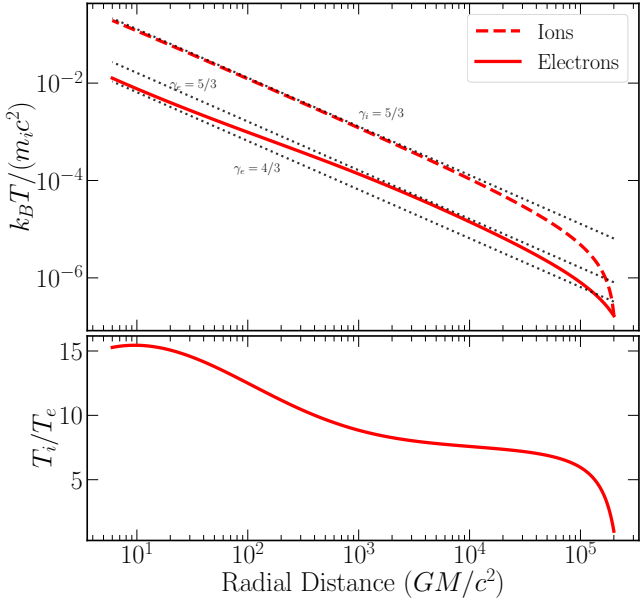}
    \caption{\textit{Top}: The ion (dashed) and electron (solid) temperatures in units of the ion rest mass energy as a function of radial distance for a flow with accretion rate $\dot{M} \ll 10^{-7} \dot{M}_{E}$, plasma $\beta = 5$, and a radial velocity profile index of $n_R = 1.5$, in the absence of the radiative cooling of the electrons. The dotted lines indicate approximate analytical solutions for different values of the adiabatic indices $\gamma$. In the inner accretion flow, where the electrons become relativistic, their temperature profile shifts towards the relativistic ($\gamma_e=4/3$) solution. \textit{Bottom}: The ion-to-electron temperature ratio as a function of radial distance for the same calculation. }
    \label{fig:turbulent heating}
\end{figure}

Figure \ref{fig:turbulent heating} shows the radial profiles of the ion and electron temperatures, as well as their ratio. As they drift inwards, both the ions and the electrons quickly heat up from their initial sub-virial temperatures at the outer radial boundary to the local virial temperatures. The ions then maintain a virial, $r^{-1}$, profile all the way to the inner region of the accretion flow. On the other hand, the electron temperature gradually shifts from one virial profile to another with a lower normalization in the inner accretion flow. 

In order to understand these solutions, it is helpful to examine an analytical limit of the solutions to equations~\eqref{eq:radial ion energy conservation} and \eqref{eq:radial electron energy conservation}, where the effects of Coulomb heating and radiative cooling are neglected and the turbulent heating rate is approximated only to leading order, assuming that it is independent of the temperature ratio $T_i/T_e$. We further simplify the equations by eliminating the temperature dependence of the effective adiabatic indices $\gamma_i$ and $\gamma_e$. In this limit, following the Appendix of \citet{2022ApJ...941...88}, we write the ion temperature as

\begin{equation}
\label{eq:ion temperature proportionality}
\dfrac{k_B T_i}{m_i c^2} = \dfrac{\gamma_i^{\rm{eff}}-1}{1-(\gamma_i^{\rm{eff}} - 1)(2-n_R)\left(1 + \dfrac{2\zeta}{\beta}\right)} \times \dfrac{f}{1+f} \times \dfrac{3}{2r},
\end{equation}
and the electron temperature as
\begin{equation}
\label{eq:electron temperature proportionality}
\dfrac{k_B T_e}{m_e c^2} = \dfrac{\gamma_e-1}{1-(\gamma_e - 1)(2-n_R)} \times \dfrac{1}{1+f} \times  \dfrac{3}{2r},
\end{equation}

\noindent where we have introduced $f=Q_i/Q_e$ as the ratio of the ion-to-electron heating rate due to the dissipation of turbulent energy. Because the ions never reach relativistic temperatures even in the inner accretion flow, their adiabatic index remains approximately $\gamma_i=5/3$ and their temperatures follows a single virial profile. However, as the electrons drift towards the black hole, their temperatures increase to relativistic values with $k_B T_e/ m_e c^2 \geq 1$, and hence their adiabatic index evolves from $\gamma_e=5/3$ in the outer regions to $\gamma_e=4/3$ inner regions of the flow. This causes their temperatures to shift from one virial profile to another that is approximately a factor of 2 lower. The temperature ratio of the ion and electrons in the outer flow, where both species are non-relativistic, is approximately equal to the ion-to-electron heating ratio, i.e., $T_i/T_e\sim Q_i/Q_e$, whereas becomes two times larger in the inner flow, where the electrons become relativistic. A similar result has been noted by \citet{Sadowski2017} based on numerical simulations. We note, however, that these authors attribute their result to the shallow density profiles of their accretion flows, which renders adiabatic heating negligible. In contrast, our analytic model demonstrates that the virial profiles of the ion and electron temperatures are practically independent of the density profile, which is controlled by the parameter $n_R$.

\begin{figure}[t]
    \includegraphics[width=0.99\linewidth]{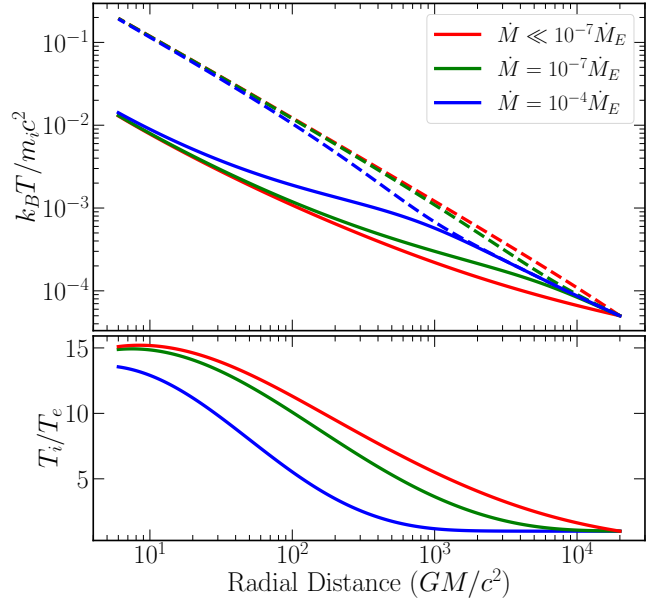}
    \caption{\textit{Top}: Ion (dashed) and electron (solid) temperatures in units of the ion rest mass energy as a function of radial distance, for  plasma $\beta = 5$ and $n_R = 1.5$, at different accretion rates, neglecting the effects of radiative cooling for the electrons. At the higher accretion rates shown, Coulomb collisions cause the ion and electron temperatures to equilibrate at large radii. \textit{Bottom}: The ion-to-electron temperature ratio, $T_i/T_e$ as a function of radial distance for the same calculation.}
    \label{fig:coulomb heating}
\end{figure}

\subsection{Coulomb Heating}

Having established the scale-free virial nature of ion and electron temperatures under turbulent heating, we now examine the effects of Coulomb coupling between ions and electrons on the flow thermodynamics. Figure \ref{fig:coulomb heating} shows the temperatures at density scales corresponding to mass accretion rates of $\dot{M}=10^{-7} \dot{M}_{E}\ \rm{and}\ 10^{-4}\dot{M}_{E}$ and also compares with the solution previously calculated with $\dot{M} \ll 10^{-7} \dot{M}_{E}$. 

At $\dot{M}=10^{-7} \dot{M}_{E}$, Coulomb collisions cause the ion and electron temperatures to equilibrate at large radii. As the collision cross section decreases with increasing temperature, the two species decouple inwards and approach their individual virial solutions set by turbulent heating. Upon further increasing the accretion rate to $\dot{M}=10^{-4} \dot{M}_{E}$, and consequently the density scale, Coulomb coupling becomes stronger. As a result, the ions and electrons remain in thermal equilibrium at even smaller radii. However, even at such accretion rates, the effects of Coulomb coupling fade away as the gas drifts inwards, with the ion and electron temperatures approaching the limit of their scale-free solutions determined by turbulent heating in the inner flow. 

\begin{figure}[t]
    \includegraphics[width=0.99\linewidth]{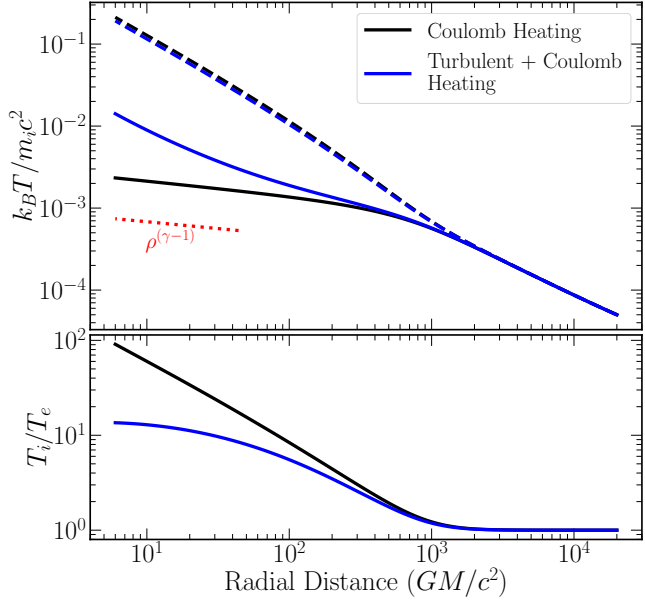}
    \caption{\textit{Top}: Ion (dashed blue) and electron (solid blue) temperatures in units of ion rest mass energy as a function of radial distance, for a mass accretion rate of $\dot{M}=10^{-4} \dot{M}_{E}$, plasma $\beta = 5$, $n_R = 1.5$, and neglecting radiative cooling effects. Black solid line shows the electron temperature profile if electrons were heated purely through Coulomb collisions with the ions. In the absence of direct heating from turbulence, the electron temperature in the inner accretion flow increases only due to adiabatic compressional heating, as shown in the red dashed line. \textit{Bottom}: The ion-to-electron temperature ratio, $T_i/T_e$ as a function of radial distance for the same calculation.}
    \label{fig:coulomb heating 2}
\end{figure}

The above analysis of the relative importance of turbulent and Coulomb heating indicates that, at density scales relevant for radiatively inefficient accretion flows, turbulent heating almost completely overwhelms the effects of Coulomb interactions in the inner regions of the accretion flow. While we will examine the effects of radiative cooling in a later subsection, it is instructive to consider certain implications of the scale-free turbulent heating of electrons. 

The initial models of radiatively inefficient accretion flows \citep[see][]{Narayan1995b} considered the electrons to be heated solely via Coulomb coupling with the ions, without any of the turbulent energy in the plasma being channeled directly into the electrons. We compare such a system with our set-up of direct electron heating in Figure~\ref{fig:coulomb heating 2}, where we show the ion and electron temperatures for a system where the electron heating is purely through Coulomb collisions with the ions. We also compare it with the direct heating model described in this paper, i.e., where a fraction of the turbulent energy is directly channeled into the electrons. The ion temperatures expectedly follow virial profiles for both these cases since the channel for ion heating remains the same. However, the electron temperatures are no longer virial, when electron heating is purely through Coulomb collisions. Instead, when the temperature increases to the point that Coulomb interactions become negligible, electrons experience pure compressional heating  ($T_e \sim n_e^{\gamma-1}$, shown in the dotted line in Figure~\ref{fig:coulomb heating 2}), which results in a significantly shallower radial profile. The consequence is an ion-to-electron temperature ratio that increases rapidly with decreasing radius to values $\sim 100$. This comparison demonstrates that, when microphysical plasma phenomena channel a finite amount of the turbulent wave energy into the electrons, this introduces substantial changes to the nature of the solution for the electron temperature compared to a setting where electrons only receive energy via Coulomb heating.

\subsection{Radiative Electron Cooling}

Having established the relative importance of particle heating via turbulent heating and Coulomb collisions, we now turn to examining the effects of electron cooling on the thermodynamics of the plasma. As described in \S \ref{sec: electron cooling subsection}, we include the effects of Bremsstrahlung and synchrotron cooling of the electrons in evolving the electron thermodynamics in the accretion disk. The plasma cooling rates due to both ion-electron and electron-electron Bremsstrahlung is proportional the square of the number density of the plasma ($\Lambda_{\rm{b}} \propto n_e^2$). On the other hand, the synchrotron cooling rate scale as $\Lambda_e\sim n_e B^2$ in our regime of interest. Because we set the local magnetic field to be proportional to the local thermal pressure using the plasma $\beta$ parametrization, the synchrotron cooling rate effectively scales with $n_e^2$, which is the same as Bremsstrahlung cooling and Coulomb heating \citep[see appendix of ][]{2022ApJ...925...13S}. 

\begin{figure}[h]
    \includegraphics[width=0.99\linewidth]{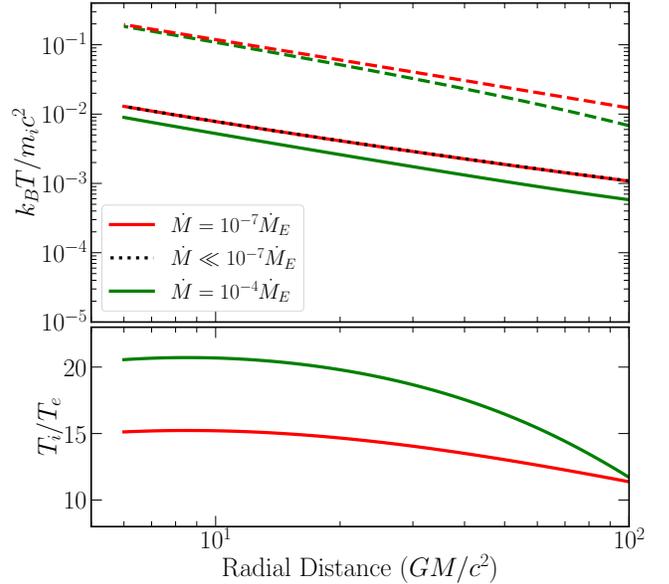}
    \caption{\textit{Top}: Ion (dashed) and Electron (solid) temperatures, in units of ion rest mass energy, calculated including the effects of radiative cooling for mass accretion rates of $\dot{M} = 10^{-7} \dot{M}_{E}$ (red), $\dot{M} \ll 10^{-7} \dot{M}_{E}$ (black dotted), and $\dot{M} = 10^{-4} \dot{M}_{E}$ (green). The plasma $\beta$ is set to $\beta = 5$ and the radial velocity profile is set by $n_R = 1.5$. \textit{Bottom}: The ion-to-electron temperature ratio, $T_i/T_e$ for the above cases.}
    \label{fig:cooling example}
\end{figure}

Figure ~\ref{fig:cooling example} shows the effects of radiative cooling processes at different accretion rates on the electron temperature and the ion-to-electron temperature ratio in the inner accretion disk. At an accretion rate of $10^{-7} \dot{M}_{E}$, radiative cooling is inefficient and the  electron temperatures are equal to the case with $\dot{M} \ll 10^{-7} \dot{M}_{E}$, for which radiative processes have been neglected. Only at higher accretion rates, such as $\dot{M} = 10^{-4}\dot{M}_{E}$, there is a small drop in the electron temperature. It is important to emphasize here, however, that because of the particular dependence of the cooling rates on the square of the density, the electron temperatures retain their virial profiles.

The effects of synchrotron cooling also depend significantly on plasma $\beta$, which sets the strength of the magnetic field. However, since the partition of turbulent heating between ions and electrons also depends strongly on plasma $\beta$, we will address the combined effect in the following subsection. 

\subsection{Heat Partition Model}
\label{sec:heating ratio solution subsection}

In the previous subsections, we established the near-virial nature of the ion and electron temperature profiles in the inner accretion flow and addressed their weak dependence on the efficiency of Coulomb collisions and radiative cooling processes. Because the ions and the electrons receive their heat mostly from turbulent dissipation in the accretion flow, the microphysical processes that partition the turbulent energy into the individual species are extremely important in determining their temperature ratios. As discussed in \S \ref{sec:partition of heating} and shown in Figure~\ref{fig:howes and kawazura model}, the partition of heat in the regime of plasma $\beta \leq 10$ depends strongly on the composition of the turbulent cascade, which is not well understood for MRI-driven turbulence. 

\begin{figure}[t]
    \includegraphics[width=0.99\linewidth]{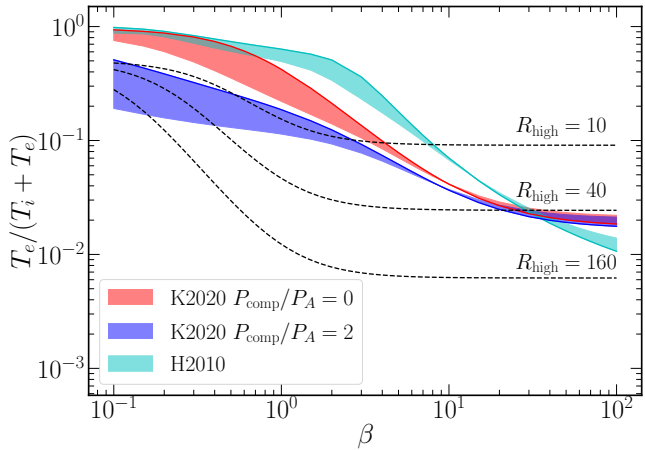}
    \caption{The ratio of the electron temperature to the total gas temperature $T_e/(T_e+T_i)$ in the inner region of the accretion flow (at $r = 10 GM/c^2$) as a function of plasma $\beta$, computed using different ratios of the compressive-to-Alfvenic power ($P_{\rm comp}/P_{A}$). The cyan curve corresponds to the earlier analytic calculation for Alfvenic turbulence by \citet{2010MNRAS.409L.104H}. The shaded region for each curve indicates the range of temperature ratios predicted for different mass accretion rates in the range $10^{-7} \le \dot{M}/\dot{M}_{\rm E} \le 10^{-5}$. In all calculations, we have set the radial velocity profile to $n_R = 1.5$. The dotted lines represent the empirical relation by \citet{Moscibrodzka2016} that has been used in the literature, for different values of the parameter $R_{\rm high}$.}
    \label{fig:heat partition ratios}
\end{figure}

In this subsection, we study the dependence of the electron temperatures on the composition of the turbulent cascades, for the regime of interest in plasma $\beta$. In particular, we will focus on the asymptotic temperature ratio $T_e/(T_e+T_i)$ in the innermost regions of the accretion flow, which measures the relative contribution of the electrons to the total gas pressure. Figure~\ref{fig:heat partition ratios}, shows this ratio calculated at $r = 10 \; GM/c^2$, as a function of plasma $\beta$, for a range of mass accretion rates of $10^{-7} \leq \dot{M}/\dot{M}_{E} \leq 10^{-5}$, and for different ratios of the compressive-to-Alfvenic wave power. 

The widths of the shaded regions in Figure~\ref{fig:heat partition ratios} represent the dependence of the temperature ratio on the accretion rate, which is determined primarily by the efficiency of synchrotron cooling. At low plasma $\beta$, where the magnetic field is stronger, synchrotron cooling is very efficient and introduces a factor of $\sim 2$ range of temperature ratios for the accretion rates we consider here, as discussed in \S\ref{sec: electron cooling subsection}. In the opposite limit of high plasma $\beta$, synchrotron cooling becomes inefficient and the temperature ratio in the inner accretion flow becomes practically independent of accretion rate.

The dependence of the temperature ratios on plasma $\beta$ simply reflects the dependence of heat partition between ions and electrons in the turbulent cascade shown in Figure~\ref{fig:howes and kawazura model}. The ratio of compressive-to-Alfvenic wave power affects primarily the temperature ratio in the low-$\beta$ regime, for which the ion-to-electron heating ratio asymptotes to $P_{\rm comp}/P_{\rm A}$, for reasons discussed earlier. In order for the turbulent cascades to heat predominantly the electrons and not the ions, i.e., leading to $T_i<T_e$, the MRI-driven turbulence would need to be purely Alfvenic.

For the latter case of purely Alfvenic turbulence, Figure~\ref{fig:heat partition ratios} also compares the temperature ratios obtained using the results of the numerical model of Alfvenic cascades (K2020 with $P_{\rm{comp}}/P_A = 0$) to those using the analytic calculation of \cite{2010MNRAS.409L.104H}. The nonlinear effects captured in the numerical calculation of the cascades lead to smaller electron temperatures, in the intermediate plasma $\beta$ range but to larger electron temperatures in the very high plasma $\beta$ regime compared to the earlier studies.

\subsection{Empirical Relations for Temperature Ratios}

In earlier calculations of the observational properties of radiatively inefficient flows based on GRMHD simulations, the ratio of the ion-to-electron temperatures have often been prescribed in post-processing, using different empirical relations that depend on plasma $\beta$~\citep[see, e.g.,][]{Chan2015,Moscibrodzka2016}. In particular, the empirical relation 
\begin{equation}
\label{eq:r high model}
R \equiv \dfrac{T_i}{T_e} = R_{\rm{high}} \dfrac{\beta^2}{1 + \beta^2} + R_{\rm{low}} \dfrac{1}{1 + \beta^2},
\end{equation}
has been used in comparing GRMHD model predictions to EHT observations \citep[see][]{2022ApJ...930L..16A}. Here $R_{\rm{high}}$ is the temperature ratio in the limit of high plasma $\beta \gg 1$, and $R_{\rm{low}}$ is the temperature ratio in the limit of low plasma $\beta \ll 1$. 

Figure~\ref{fig:heat partition ratios} compares this empirical relation to our calculation of the temperature ratio. For the empirical ratio, we have set $R_{\rm{low}}=1$, as has been typically adopted. We find that, at the high plasma $\beta$ limit, the asymptotic temperature ratio corresponds to $R_{\rm high}\simeq 40$. However, the functional dependence of the temperature ratio on plasma $\beta$ is not captured by empirical relation that is typically employed. Moreover, the low plasma $\beta$ limit given by $R_{\rm low}$ does not capture the dependence of the temperature ratio on the composition of the turbulent cascade. 

Our model of the accretion flow allows us to generate a new, physically motivated empirical relation for the ion-to-electron temperature ratio. Taking the ratio of equations~(\ref{eq:ion temperature proportionality}) and (\ref{eq:electron temperature proportionality}), we obtain
\begin{equation}
\label{eq:empirical}
    \dfrac{T_i}{T_e} = \left(\dfrac{\gamma_i^{\rm{eff}}-1}{\gamma_e-1} \right) \dfrac{1 - (\gamma_e-1)(2-n_R)}{1 - (\gamma_i^{\rm{eff}}-1)(2-n_R)\left(1 + \dfrac{2\zeta}{\beta}\right)} \left(\frac{Q_i}{Q_e}\right)\;.
\end{equation}
Suppressing the weak dependence on temperature of the partition of heat between ion and electrons in equation~(\ref{eq:kawazura model}), we write
\begin{equation}
\label{eq:kawazura_reduced}
\frac{Q_i}{Q_e} = \dfrac{35}{1 + (\beta/15)^{-1.4}} + \dfrac{P_{\rm{comp}}}{P_{\rm{AW}}}. 
\end{equation}
We estimate the effective adiabatic index of the ions using equation~(\ref{eq: internal energy ions}) to write
\begin{equation}
\dfrac{1}{\gamma_i^{\rm{eff}}-1}=\left( \frac{1}{\gamma_i - 1} \right) \left(1 + \dfrac{2\zeta}{\beta} \right).
\end{equation} 
Because the ions remain subrelativistic throughout the flow, we set $\gamma_i=5/3$. 

As discussed above, the temperature of the electrons and, hence, their adiabatic index, depends on the value of plasma $\beta$. At low values of plasma $\beta$, the electrons get to sufficiently high tempeartures to become ultra relativistic. In this limit, their adiabatic index is $\lim_{\beta\rightarrow 0}\gamma_e=4/3$. In the regime of high plasma $\beta$, the electron temperature is significantly lower and their adiabatic index increases towards the non-relativistic value of 5/3. In principle, one could use expression~(\ref{eq:gamma trans relativistic}) for the temperature dependence of the adiabatic index and solve the implicit equation for the electron temperature. For computational efficiency, we have devised an empirical relation for the adiabatic index of the electrons in the inner accretion flow using the form
\begin{equation}
    \label{eq:gamma e eff}
    \gamma_e = \frac{4}{3} + \kappa \dfrac{(\beta/\beta_{\rm{break}})^2}{1 + (\beta/\beta_{\rm{break}})^2}.
\end{equation}

We have then set $\kappa = 0.13$ and $\beta_{\rm{break}} = 5$ in order to reproduced the behavior seen in the detailed solutions of our accretion model.

\begin{figure}[t]
    \centering
    \includegraphics[width=0.99\linewidth]{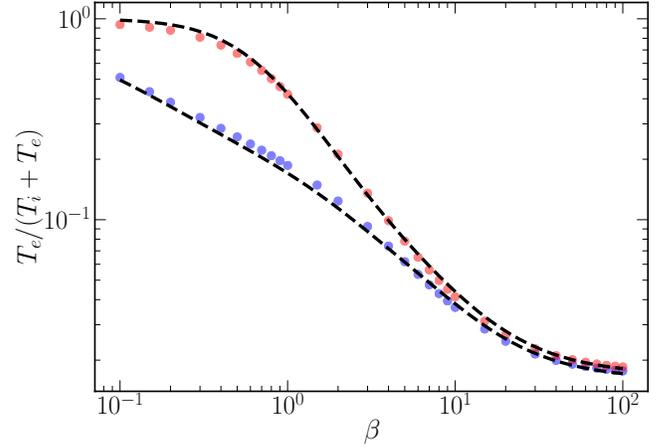}
    \caption{The ratio of the electron temperature to the total gas temperature for $\dot{M}\le 10^{-7} \dot{M}_{\rm E}$ in the inner region of a radiatively inefficient accretion flow ($r = 10 GM/c^2$), as a function of plasma $\beta$, calculated using our covariant, semi-analytic model, for different ratios of the compressive-to-Alfvenic power ($P_{\rm comp}/P_{\rm A}$). The dashed lines correspond to the empirical relation~\eqref{eq:empirical}.}
    \label{fig:empirical_relation}
\end{figure}

Figure~\ref{fig:empirical_relation} compares the ratio of the electron-to-total gas temperatures calculated using our covariant semi-analytical model of the accretion flow to the empirical relation~(\ref{eq:empirical}), using expressions \eqref{eq:kawazura_reduced}--\eqref{eq:gamma e eff} for its parameters. For this comparison, we have set the velocity profile parameter to $n_R=1.5$ and neglected the effects of radiative cooling. The latter are expected to be relatively unimportant for the accretion flows around the two prime imaging targets of the EHT, Sgr~A* and M87.

The empirical relation~(\ref{eq:empirical}) can be easily implemented in a calculation of observables based on postprocessing of GRMHD simulations. The $n_R$ velocity profile parameter can be inferred empirically from the time- and azimuthally-averaged velocity (or density) profile in the simulation.  Finally, the ratio of the compressive-to-Alfvenic wave power can be either set to the value inferred by \cite{2022JPlPh..88c9011K} or allowed to be a free parameter to be inferred observationally.

\section{Discussion}
\label{sec:Discussion}

Recent studies of rarefied plasmas, such as those that exist in low-luminosity accretion flows, have uncovered particle heating channels that affect their thermodynamic properties. In particular, turbulent cascades of compressive and Alfvenic waves channel heat directly into ions and electrons, with the partition depending on the ratio of power in these waves and the plasma $\beta$ \citep{2020PhRvX..10d1050K}.

In this paper, we explored the effect of turbulent heating on the temperature profiles of ions and electrons using a covariant analytic model. We implemented the effects of compressive and Alfvenic dissipation, the exchange of energy between the two species through Coulomb collisions, compressional heating, as well as radiative cooling for the electrons. We demonstrated that both the electrons and the ions follow nearly parallel radial profiles with a temperature ratio that is determined by three properties of the plasma: the $\beta$ parameter, the ratio of turbulent energy residing in compressive and Alfvenic cascades, the accretion rate that sets the efficiency of radiative cooling. In the inner accretion flows for the conditions that are found in radiatively inefficient accretion flows, this ratio ranges between $\sim 5$ and $\sim 40$ when these processes are considered. 

There exist some additional dissipative processes not considered in the present work that can differentially channel energy into the electrons or the ions and, therefore, impact their temperature profiles and ratio. Magnetic reconnection injects energy into both species \citep{2016ApJ...826...77B,2018ApJ...853..184B,2017ApJ...850...29R,2019ApJ...873....2R}, but the partition of heat $Q_i/Q_e$ can increase or decrease with plasma $\beta$ depending on the strength of the guide field \citep{2019ApJ...873....2R}. Moreover, reconnection can accelerate electrons to large Lorentz factors well above the thermal distribution. The impact of this channel in radiatively inefficient flows depends on the fraction of internal energy that is dissipated via magnetic reconnection, which is not known. A second dissipative process arises from the presence of velocity-space instabilities in weakly collisional plasmas \citep{2015ApJ...800...88S,2007ApJ...667..714S}. In this channel, the partition can differ from the pressure-isotropic case for plasmas that are near the threshold of mirror and firehose instabilities. Pressure anisotropies can also cause dissipation at larger scales and impact the partition of viscous heat among the species \citep{2019MNRAS.486.4013K,2018JPlPh..84b7101K}. The relevance of these processes for plasmas in radiatively inefficient flows needs to be further explored. 

The temperature profiles in the plasma, especially those of the electrons, connect directly to the observable properties of accreting black holes, such as their spectra, variability, and images. Our calculations allow us to devise an empirical relation for the ion-to-electron temperatures that depend not only on all plasma parameters known to be relevant for particle heating but also and on the global flow properties that determine compressional heating. This empirical relation is easy to implement into the outputs of global GRMHD simulations to calculate observational signatures. 

A key parameter that controls the ion-to-electron temperature ratio is the partition of turbulent energy injected into compressive and Alfvenic cascades. 
\citet{2022JPlPh..88c9011K} provided a first estimate of the composition of a turbulent cascade in an accretion flow driven by the MRI in the presence of a strong azimuthal magnetic field, which we use in our calculations. Further  simulations are needed to quantify this partition under a broader range of conditions. 

As observations of radiatively inefficient flows have evolved from primarily acquiring spectral data \citep[see][for a review]{2014ARA&A..52..529Y} to horizon-scale imaging and polarimetric observations with very long baseline interferometry \citep{2019ApJ...875L...1E,2021ApJ...910L..12E,2022ApJ...930L..12A}, it is becoming increasingly possible to probe plasma processes directly. Improved simulations that incorporate the new physical effects will allow us to make more predictive models of black hole environments. Conversely, we can use the observations that directly probe the densities, temperatures, and magnetic fields of accretion flows to guide and constrain the studies of microphysics of turbulent rarefied plasmas \citep[see, e.g.,][]{Xie2023}.

\acknowledgements
The authors acknowledge support from the NASA ATP award 80NSSC20K0521 and NSF PIRE award OISE-1743747. We thank K.\,G.\, Klein and C.-K.\, Chan for useful discussions. 

\appendix

\section{General Relativistic and Inner Boundary Effects}

In this appendix, we examine the role of general relativistic effects related to the black-hole spin and the viscous stresses at the innermost stable circular orbit (ISCO) in determining the global temperature profile of the plasma species. Our model for the radial velocity of the plasma only considers the centrifugally supported flow outside the ISCO (eq.~\eqref{eq:radial plasma velocity}) and is not influenced by the black-hole spin. The only quantity affected by the spin is the location of the ISCO, whereas the plasma heating rate, $Q$, depends on both the location of the ISCO and the viscous stresses there (eq.~\eqref{eq:expression for Q}).   

\begin{figure}[t]
    \centerline{\includegraphics[width=0.5\linewidth]{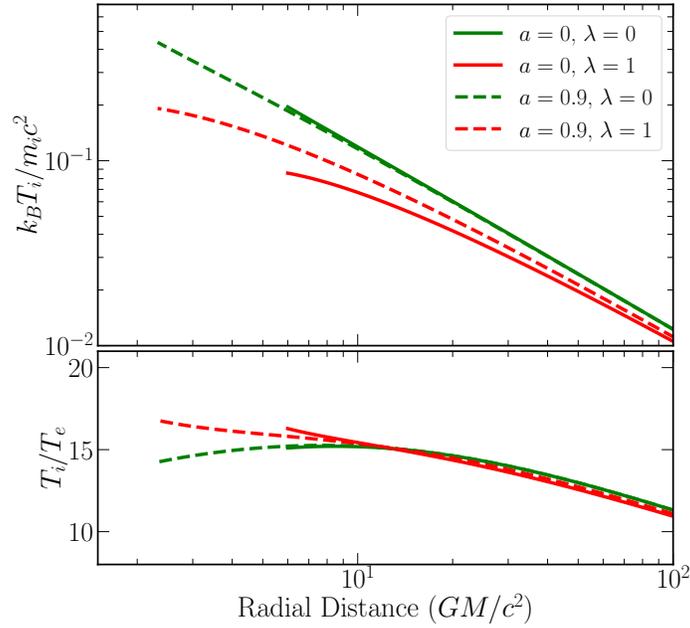}}
    \caption{\textit{Top}: The ion temperature in units of ion rest mass energy as a function of radial distance, computed for a non-spinning black-hole (dashed lines), and a black-hole with a spin $a = 0.9$ (solid lines). The lines shown in green correspond to flows with vanishing viscous stresses at the ISCO ($\lambda = 1$), while the opposite extreme corresponds to $\lambda=0$. In all calculations, we have set the plasma parameter to $\beta=5$, the radial velocity profile to $n_R=1.5$, and the mass accretion rate set to $\dot{M} \ll 10^{-7}\dot{M}_{E}$. \textit{Bottom}: The ion-to-electron temperature ratio, $T_i/T_e$.}
    \label{fig:effects of GR and viscous stresses}
\end{figure}

Figure \ref{fig:effects of GR and viscous stresses} shows the ion temperatures and the ion-to-electron temperature ratios near the ISCO, for different values of the black-hole spin and the parameter $\lambda$ that sets the stresses at the ISCO; all other model parameters are fixed at their fiducial values, i.e., plasma $\beta = 5$ and $n_R = 1.5$. We also ignore the contributions of electron cooling and Coulomb heating, which are negligible in the inner accretion flow.

As the specific angular momentum of the flow at the ISCO increases towards its local Keplerian value, stresses and the corresponding turbulent heating rate of the plasma decreases substantially, (see eq.~\eqref{eq:expression for Q}). This causes a drop in the overall temperature of the ions by a factor of $\sim 3$. However, the effect on the ion-to-electron temperature ratio of changing the black-hole spin or the stresses at the ISCO is only marginal.

\bibliographystyle{apj}
\bibliography{salient}

\begin{thebibliography}{}
\expandafter\ifx\csname natexlab\endcsname\relax\def\natexlab#1{#1}\fi

\bibitem[{{Akiyama} {et~al.}(2022{\natexlab{a}}){Akiyama}, {Alberdi}, {Alef},
  {Algaba}, {Anantua}, {Asada}, {Azulay}, {Bach}, {Baczko}, {Ball},
  {Balokovi{\'c}}, {Barrett}, {Baub{\"o}ck}, {Benson}, {Bintley}, {Blackburn},
  {Blundell}, {Bouman}, {Bower}, {Boyce}, {Bremer}, {Brinkerink}, {Brissenden},
  {Britzen}, {Broderick}, {Broguiere}, {Bronzwaer}, {Bustamante}, {Byun},
  {Carlstrom}, {Ceccobello}, {Chael}, {Chan}, {Chatterjee}, {Chatterjee},
  {Chen}, {Chen}, {Cheng}, {Cho}, {Christian}, {Conroy}, {Conway}, {Cordes},
  {Crawford}, {Crew}, {Cruz-Osorio}, {Cui}, {Davelaar}, {De Laurentis},
  {Deane}, {Dempsey}, {Desvignes}, {Dexter}, {Dhruv}, {Doeleman}, {Dougal},
  {Dzib}, {Eatough}, {Emami}, {Falcke}, {Farah}, {Fish}, {Fomalont}, {Ford},
  {Fraga-Encinas}, {Freeman}, {Friberg}, {Fromm}, {Fuentes}, {Galison},
  {Gammie}, {Garc{\'\i}a}, {Gentaz}, {Georgiev}, {Goddi}, {Gold},
  {G{\'o}mez-Ruiz}, {G{\'o}mez}, {Gu}, {Gurwell}, {Hada}, {Haggard}, {Haworth},
  {Hecht}, {Hesper}, {Heumann}, {Ho}, {Ho}, {Honma}, {Huang}, {Huang},
  {Hughes}, {Ikeda}, {Impellizzeri}, {Inoue}, {Issaoun}, {James}, {Jannuzi},
  {Janssen}, {Jeter}, {Jiang}, {Jim{\'e}nez-Rosales}, {Johnson}, {Jorstad},
  {Joshi}, {Jung}, {Karami}, {Karuppusamy}, {Kawashima}, {Keating}, {Kettenis},
  {Kim}, {Kim}, {Kim}, {Kim}, {Kino}, {Koay}, {Kocherlakota}, {Kofuji}, {Koch},
  {Koyama}, {Kramer}, {Kramer}, {Krichbaum}, {Kuo}, {La Bella}, {Lauer}, {Lee},
  {Lee}, {Leung}, {Levis}, {Li}, {Lico}, {Lindahl}, {Lindqvist}, {Lisakov},
  {Liu}, {Liu}, {Liuzzo}, {Lo}, {Lobanov}, {Loinard}, {Lonsdale}, {Lu}, {Mao},
  {Marchili}, {Markoff}, {Marrone}, {Marscher}, {Mart{\'\i}-Vidal},
  {Matsushita}, {Matthews}, {Medeiros}, {Menten}, {Michalik}, {Mizuno},
  {Mizuno}, {Moran}, {Moriyama}, {Moscibrodzka}, {M{\"u}ller}, {Mus}, {Musoke},
  {Myserlis}, {Nadolski}, {Nagai}, {Nagar}, {Nakamura}, {Narayan}, {Narayanan},
  {Natarajan}, {Nathanail}, {Navarro Fuentes}, {Neilsen}, {Neri}, {Ni},
  {Noutsos}, {Nowak}, {Oh}, {Okino}, {Olivares}, {Ortiz-Le{\'o}n}, {Oyama},
  {{\"O}zel}, {Palumbo}, {Paraschos}, {Park}, {Parsons}, {Patel}, {Pen},
  {Pesce}, {Pi{\'e}tu}, {Plambeck}, {PopStefanija}, {Porth}, {P{\"o}tzl},
  {Prather}, {Preciado-L{\'o}pez}, {Psaltis}, {Pu}, {Ramakrishnan}, {Rao},
  {Rawlings}, {Raymond}, {Rezzolla}, {Ricarte}, {Ripperda}, {Roelofs},
  {Rogers}, {Ros}, {Romero-Ca{\~n}izales}, {Roshanineshat}, {Rottmann}, {Roy},
  {Ruiz}, {Ruszczyk}, {Rygl}, {S{\'a}nchez}, {S{\'a}nchez-Arg{\"u}elles},
  {S{\'a}nchez-Portal}, {Sasada}, {Satapathy}, {Savolainen}, {Schloerb},
  {Schonfeld}, {Schuster}, {Shao}, {Shen}, {Small}, {Sohn}, {SooHoo},
  {Souccar}, {Sun}, {Tazaki}, {Tetarenko}, {Tiede}, {Tilanus}, {Titus},
  {Torne}, {Traianou}, {Trent}, {Trippe}, {Turk}, {van Bemmel}, {van
  Langevelde}, {van Rossum}, {Vos}, {Wagner}, {Ward-Thompson}, {Wardle},
  {Weintroub}, {Wex}, {Wharton}, {Wielgus}, {Wiik}, {Witzel}, {Wondrak},
  {Wong}, {Wu}, {Yamaguchi}, {Yoon}, {Young}, {Young}, {Younsi}, {Yuan},
  {Yuan}, {Zensus}, {Zhang}, {Zhao}, {Zhao}, {Agurto}, {Allardi}, {Amestica},
  {Araneda}, {Arriagada}, {Berghuis}, {Bertarini}, {Berthold}, {Blanchard},
  {Brown}, {C{\'a}rdenas}, {Cantzler}, {Caro}, {Castillo-Dom{\'\i}nguez},
  {Chan}, {Chang}, {Chang}, {Chang}, {Chang}, {Chen}, {Chilson}, {Chuter},
  {Ciechanowicz}, {Colin-Beltran}, {Coulson}, {Crowley}, {Degenaar},
  {Dornbusch}, {Dur{\'a}n}, {Everett}, {Faber}, {Forster}, {Fuchs}, {Gale},
  {Geertsema}, {Gonz{\'a}lez}, {Graham}, {Gueth}, {Halverson}, {Han}, {Han},
  {Hasegawa}, {Hern{\'a}ndez-Rebollar}, {Herrera}, {Herrero-Illana},
  {Heyminck}, {Hirota}, {Hoge}, {Hostler Schimpf}, {Howie}, {Huang}, {Jiang},
  {Jinchi}, {John}, {Kimura}, {Klein}, {Kubo}, {Kuroda}, {Kwon}, {Lacasse},
  {Laing}, {Leitch}, {Li}, {Liu}, {Liu}, {Lin}, {Lu}, {Mac-Auliffe},
  {Martin-Cocher}, {Matulonis}, {Maute}, {Messias}, {Meyer-Zhao},
  {Monta{\~n}a}, {Montenegro-Montes}, {Montgomerie}, {Moreno Nolasco},
  {Muders}, {Nishioka}, {Norton}, {Nystrom}, {Ogawa}, {Olivares}, {Oshiro},
  {P{\'e}rez-Beaupuits}, {Parra}, {Phillips}, {Poirier}, {Pradel}, {Qiu},
  {Raffin}, {Rahlin}, {Ram{\'\i}rez}, {Ressler}, {Reynolds},
  {Rodr{\'\i}guez-Montoya}, {Saez-Madain}, {Santana}, {Shaw}, {Shirkey},
  {Silva}, {Snow}, {Sousa}, {Sridharan}, {Stahm}, {Stark}, {Test},
  {Torstensson}, {Venegas}, {Walther}, {Wei}, {White}, {Wieching}, {Wijnands},
  {Wouterloot}, {Yu}, {Yu}, {Zeballos}, \& {EHT
  Collaboration}}]{2022ApJ...930L..12A}
{Akiyama}, K., {Alberdi}, A., {Alef}, W., {et~al.} 2022{\natexlab{a}}, \apjl,
  930, L12

\bibitem[{{Akiyama} {et~al.}(2022{\natexlab{b}}){Akiyama}, {Alberdi}, {Alef},
  {Algaba}, {Anantua}, {Asada}, {Azulay}, {Bach}, {Baczko}, {Ball},
  {Balokovi{\'c}}, {Barrett}, {Baub{\"o}ck}, {Benson}, {Bintley}, {Blackburn},
  {Blundell}, {Bouman}, {Bower}, {Boyce}, {Bremer}, {Brinkerink}, {Brissenden},
  {Britzen}, {Broderick}, {Broguiere}, {Bronzwaer}, {Bustamante}, {Byun},
  {Carlstrom}, {Ceccobello}, {Chael}, {Chan}, {Chatterjee}, {Chatterjee},
  {Chen}, {Chen}, {Cheng}, {Cho}, {Christian}, {Conroy}, {Conway}, {Cordes},
  {Crawford}, {Crew}, {Cruz-Osorio}, {Cui}, {Davelaar}, {De Laurentis},
  {Deane}, {Dempsey}, {Desvignes}, {Dexter}, {Dhruv}, {Doeleman}, {Dougal},
  {Dzib}, {Eatough}, {Emami}, {Falcke}, {Farah}, {Fish}, {Fomalont}, {Ford},
  {Fraga-Encinas}, {Freeman}, {Friberg}, {Fromm}, {Fuentes}, {Galison},
  {Gammie}, {Garc{\'\i}a}, {Gentaz}, {Georgiev}, {Goddi}, {Gold},
  {G{\'o}mez-Ruiz}, {G{\'o}mez}, {Gu}, {Gurwell}, {Hada}, {Haggard}, {Haworth},
  {Hecht}, {Hesper}, {Heumann}, {Ho}, {Ho}, {Honma}, {Huang}, {Huang},
  {Hughes}, {Ikeda}, {Impellizzeri}, {Inoue}, {Issaoun}, {James}, {Jannuzi},
  {Janssen}, {Jeter}, {Jiang}, {Jim{\'e}nez-Rosales}, {Johnson}, {Jorstad},
  {Joshi}, {Jung}, {Karami}, {Karuppusamy}, {Kawashima}, {Keating}, {Kettenis},
  {Kim}, {Kim}, {Kim}, {Kim}, {Kino}, {Koay}, {Kocherlakota}, {Kofuji}, {Koch},
  {Koyama}, {Kramer}, {Kramer}, {Krichbaum}, {Kuo}, {La Bella}, {Lauer}, {Lee},
  {Lee}, {Leung}, {Levis}, {Li}, {Lico}, {Lindahl}, {Lindqvist}, {Lisakov},
  {Liu}, {Liu}, {Liuzzo}, {Lo}, {Lobanov}, {Loinard}, {Lonsdale}, {Lu}, {Mao},
  {Marchili}, {Markoff}, {Marrone}, {Marscher}, {Mart{\'\i}-Vidal},
  {Matsushita}, {Matthews}, {Medeiros}, {Menten}, {Michalik}, {Mizuno},
  {Mizuno}, {Moran}, {Moriyama}, {Moscibrodzka}, {M{\"u}ller}, {Mus}, {Musoke},
  {Myserlis}, {Nadolski}, {Nagai}, {Nagar}, {Nakamura}, {Narayan}, {Narayanan},
  {Natarajan}, {Nathanail}, {Navarro Fuentes}, {Neilsen}, {Neri}, {Ni},
  {Noutsos}, {Nowak}, {Oh}, {Okino}, {Olivares}, {Ortiz-Le{\'o}n}, {Oyama},
  {{\"O}zel}, {Palumbo}, {Paraschos}, {Park}, {Parsons}, {Patel}, {Pen},
  {Pesce}, {Pi{\'e}tu}, {Plambeck}, {PopStefanija}, {Porth}, {P{\"o}tzl},
  {Prather}, {Preciado-L{\'o}pez}, {Psaltis}, {Pu}, {Ramakrishnan}, {Rao},
  {Rawlings}, {Raymond}, {Rezzolla}, {Ricarte}, {Ripperda}, {Roelofs},
  {Rogers}, {Ros}, {Romero-Ca{\~n}izales}, {Roshanineshat}, {Rottmann}, {Roy},
  {Ruiz}, {Ruszczyk}, {Rygl}, {S{\'a}nchez}, {S{\'a}nchez-Arg{\"u}elles},
  {S{\'a}nchez-Portal}, {Sasada}, {Satapathy}, {Savolainen}, {Schloerb},
  {Schonfeld}, {Schuster}, {Shao}, {Shen}, {Small}, {Sohn}, {SooHoo},
  {Souccar}, {Sun}, {Tazaki}, {Tetarenko}, {Tiede}, {Tilanus}, {Titus},
  {Torne}, {Traianou}, {Trent}, {Trippe}, {Turk}, {van Bemmel}, {van
  Langevelde}, {van Rossum}, {Vos}, {Wagner}, {Ward-Thompson}, {Wardle},
  {Weintroub}, {Wex}, {Wharton}, {Wielgus}, {Wiik}, {Witzel}, {Wondrak},
  {Wong}, {Wu}, {Yamaguchi}, {Yoon}, {Young}, {Young}, {Younsi}, {Yuan},
  {Yuan}, {Zensus}, {Zhang}, {Zhao}, {Zhao}, {Chan}, {Qiu}, {Ressler}, {White},
  \& {EHT Collaboration}}]{2022ApJ...930L..16A}
---. 2022{\natexlab{b}}, \apjl, 930, L16

\bibitem[{{Balbus} \& {Hawley}(1991)}]{1991ApJ...376..214B}
{Balbus}, S.~A., \& {Hawley}, J.~F. 1991, \apj, 376, 214

\bibitem[{{Ball} {et~al.}(2016){Ball}, {{\"O}zel}, {Psaltis}, \&
  {Chan}}]{2016ApJ...826...77B}
{Ball}, D., {{\"O}zel}, F., {Psaltis}, D., \& {Chan}, C.-k. 2016, \apj, 826, 77

\bibitem[{{Ball} {et~al.}(2018){Ball}, {{\"O}zel}, {Psaltis}, {Chan}, \&
  {Sironi}}]{2018ApJ...853..184B}
{Ball}, D., {{\"O}zel}, F., {Psaltis}, D., {Chan}, C.-K., \& {Sironi}, L. 2018,
  \apj, 853, 184

\bibitem[{{Chael} {et~al.}(2018){Chael}, {Rowan}, {Narayan}, {Johnson}, \&
  {Sironi}}]{2018MNRAS.478.5209C}
{Chael}, A., {Rowan}, M., {Narayan}, R., {Johnson}, M., \& {Sironi}, L. 2018,
  \mnras, 478, 5209

\bibitem[{{Chan} {et~al.}(2015){Chan}, {Psaltis}, {{\"O}zel}, {Narayan}, \&
  {S{\k{a}}dowski}}]{Chan2015}
{Chan}, C.-k., {Psaltis}, D., {{\"O}zel}, F., {Narayan}, R., \&
  {S{\k{a}}dowski}, A. 2015, \apj, 799, 1

\bibitem[{{Colpi} {et~al.}(1984){Colpi}, {Maraschi}, \&
  {Treves}}]{1984ApJ...280..319C}
{Colpi}, M., {Maraschi}, L., \& {Treves}, A. 1984, \apj, 280, 319

\bibitem[{{De Villiers} \& {Hawley}(2003)}]{deVilliers2003}
{De Villiers}, J.-P., \& {Hawley}, J.~F. 2003, \apj, 592, 1060

\bibitem[{{Event Horizon Telescope Collaboration}
  {et~al.}(2019{\natexlab{a}}){Event Horizon Telescope Collaboration},
  {Akiyama}, {Alberdi}, {Alef}, {Asada}, {Azulay}, {Baczko}, {Ball},
  {Balokovi{\'c}}, {Barrett}, {Bintley}, {Blackburn}, {Boland}, {Bouman},
  {Bower}, {Bremer}, {Brinkerink}, {Brissenden}, {Britzen}, {Broderick},
  {Broguiere}, {Bronzwaer}, {Byun}, {Carlstrom}, {Chael}, {Chan}, {Chatterjee},
  {Chatterjee}, {Chen}, {Chen}, {Cho}, {Christian}, {Conway}, {Cordes}, {Crew},
  {Cui}, {Davelaar}, {De Laurentis}, {Deane}, {Dempsey}, {Desvignes}, {Dexter},
  {Doeleman}, {Eatough}, {Falcke}, {Fish}, {Fomalont}, {Fraga-Encinas},
  {Freeman}, {Friberg}, {Fromm}, {G{\'o}mez}, {Galison}, {Gammie},
  {Garc{\'\i}a}, {Gentaz}, {Georgiev}, {Goddi}, {Gold}, {Gu}, {Gurwell},
  {Hada}, {Hecht}, {Hesper}, {Ho}, {Ho}, {Honma}, {Huang}, {Huang}, {Hughes},
  {Ikeda}, {Inoue}, {Issaoun}, {James}, {Jannuzi}, {Janssen}, {Jeter}, {Jiang},
  {Johnson}, {Jorstad}, {Jung}, {Karami}, {Karuppusamy}, {Kawashima},
  {Keating}, {Kettenis}, {Kim}, {Kim}, {Kim}, {Kino}, {Koay}, {Koch}, {Koyama},
  {Kramer}, {Kramer}, {Krichbaum}, {Kuo}, {Lauer}, {Lee}, {Li}, {Li},
  {Lindqvist}, {Liu}, {Liuzzo}, {Lo}, {Lobanov}, {Loinard}, {Lonsdale}, {Lu},
  {MacDonald}, {Mao}, {Markoff}, {Marrone}, {Marscher}, {Mart{\'\i}-Vidal},
  {Matsushita}, {Matthews}, {Medeiros}, {Menten}, {Mizuno}, {Mizuno}, {Moran},
  {Moriyama}, {Moscibrodzka}, {M{\"u}ller}, {Nagai}, {Nagar}, {Nakamura},
  {Narayan}, {Narayanan}, {Natarajan}, {Neri}, {Ni}, {Noutsos}, {Okino},
  {Olivares}, {Ortiz-Le{\'o}n}, {Oyama}, {{\"O}zel}, {Palumbo}, {Patel}, {Pen},
  {Pesce}, {Pi{\'e}tu}, {Plambeck}, {PopStefanija}, {Porth}, {Prather},
  {Preciado-L{\'o}pez}, {Psaltis}, {Pu}, {Ramakrishnan}, {Rao}, {Rawlings},
  {Raymond}, {Rezzolla}, {Ripperda}, {Roelofs}, {Rogers}, {Ros}, {Rose},
  {Roshanineshat}, {Rottmann}, {Roy}, {Ruszczyk}, {Ryan}, {Rygl},
  {S{\'a}nchez}, {S{\'a}nchez-Arguelles}, {Sasada}, {Savolainen}, {Schloerb},
  {Schuster}, {Shao}, {Shen}, {Small}, {Sohn}, {SooHoo}, {Tazaki}, {Tiede},
  {Tilanus}, {Titus}, {Toma}, {Torne}, {Trent}, {Trippe}, {Tsuda}, {van
  Bemmel}, {van Langevelde}, {van Rossum}, {Wagner}, {Wardle}, {Weintroub},
  {Wex}, {Wharton}, {Wielgus}, {Wong}, {Wu}, {Young}, {Young}, {Younsi},
  {Yuan}, {Yuan}, {Zensus}, {Zhao}, {Zhao}, {Zhu}, {Algaba}, {Allardi},
  {Amestica}, {Anczarski}, {Bach}, {Baganoff}, {Beaudoin}, {Benson},
  {Berthold}, {Blanchard}, {Blundell}, {Bustamente}, {Cappallo},
  {Castillo-Dom{\'\i}nguez}, {Chang}, {Chang}, {Chang}, {Chen}, {Chilson},
  {Chuter}, {C{\'o}rdova Rosado}, {Coulson}, {Crawford}, {Crowley}, {David},
  {Derome}, {Dexter}, {Dornbusch}, {Dudevoir}, {Dzib}, {Eckart}, {Eckert},
  {Erickson}, {Everett}, {Faber}, {Farah}, {Fath}, {Folkers}, {Forbes},
  {Freund}, {G{\'o}mez-Ruiz}, {Gale}, {Gao}, {Geertsema}, {Graham}, {Greer},
  {Grosslein}, {Gueth}, {Haggard}, {Halverson}, {Han}, {Han}, {Hao},
  {Hasegawa}, {Henning}, {Hern{\'a}ndez-G{\'o}mez}, {Herrero-Illana},
  {Heyminck}, {Hirota}, {Hoge}, {Huang}, {Impellizzeri}, {Jiang}, {Kamble},
  {Keisler}, {Kimura}, {Kono}, {Kubo}, {Kuroda}, {Lacasse}, {Laing}, {Leitch},
  {Li}, {Lin}, {Liu}, {Liu}, {Lu}, {Marson}, {Martin-Cocher}, {Massingill},
  {Matulonis}, {McColl}, {McWhirter}, {Messias}, {Meyer-Zhao}, {Michalik},
  {Monta{\~n}a}, {Montgomerie}, {Mora-Klein}, {Muders}, {Nadolski}, {Navarro},
  {Neilsen}, {Nguyen}, {Nishioka}, {Norton}, {Nowak}, {Nystrom}, {Ogawa},
  {Oshiro}, {Oyama}, {Parsons}, {Paine}, {Pe{\~n}alver}, {Phillips}, {Poirier},
  {Pradel}, {Primiani}, {Raffin}, {Rahlin}, {Reiland}, {Risacher}, {Ruiz},
  {S{\'a}ez-Mada{\'\i}n}, {Sassella}, {Schellart}, {Shaw}, {Silva}, {Shiokawa},
  {Smith}, {Snow}, {Souccar}, {Sousa}, {Sridharan}, {Srinivasan}, {Stahm},
  {Stark}, {Story}, {Timmer}, {Vertatschitsch}, {Walther}, {Wei}, {Whitehorn},
  {Whitney}, {Woody}, {Wouterloot}, {Wright}, {Yamaguchi}, {Yu}, {Zeballos},
  {Zhang}, \& {Ziurys}}]{2019ApJ...875L...1E}
{Event Horizon Telescope Collaboration}, {Akiyama}, K., {Alberdi}, A., {et~al.}
  2019{\natexlab{a}}, \apjl, 875, L1

\bibitem[{{Event Horizon Telescope Collaboration}
  {et~al.}(2019{\natexlab{b}}){Event Horizon Telescope Collaboration},
  {Akiyama}, {Alberdi}, {Alef}, {Asada}, {Azulay}, {Baczko}, {Ball},
  {Balokovi{\'c}}, {Barrett}, {Bintley}, {Blackburn}, {Boland}, {Bouman},
  {Bower}, {Bremer}, {Brinkerink}, {Brissenden}, {Britzen}, {Broderick},
  {Broguiere}, {Bronzwaer}, {Byun}, {Carlstrom}, {Chael}, {Chan}, {Chatterjee},
  {Chatterjee}, {Chen}, {Chen}, {Cho}, {Christian}, {Conway}, {Cordes}, {Crew},
  {Cui}, {Davelaar}, {De Laurentis}, {Deane}, {Dempsey}, {Desvignes}, {Dexter},
  {Doeleman}, {Eatough}, {Falcke}, {Fish}, {Fomalont}, {Fraga-Encinas},
  {Friberg}, {Fromm}, {G{\'o}mez}, {Galison}, {Gammie}, {Garc{\'\i}a},
  {Gentaz}, {Georgiev}, {Goddi}, {Gold}, {Gu}, {Gurwell}, {Hada}, {Hecht},
  {Hesper}, {Ho}, {Ho}, {Honma}, {Huang}, {Huang}, {Hughes}, {Ikeda}, {Inoue},
  {Issaoun}, {James}, {Jannuzi}, {Janssen}, {Jeter}, {Jiang}, {Johnson},
  {Jorstad}, {Jung}, {Karami}, {Karuppusamy}, {Kawashima}, {Keating},
  {Kettenis}, {Kim}, {Kim}, {Kim}, {Kino}, {Koay}, {Koch}, {Koyama}, {Kramer},
  {Kramer}, {Krichbaum}, {Kuo}, {Lauer}, {Lee}, {Li}, {Li}, {Lindqvist}, {Liu},
  {Liuzzo}, {Lo}, {Lobanov}, {Loinard}, {Lonsdale}, {Lu}, {MacDonald}, {Mao},
  {Markoff}, {Marrone}, {Marscher}, {Mart{\'\i}-Vidal}, {Matsushita},
  {Matthews}, {Medeiros}, {Menten}, {Mizuno}, {Mizuno}, {Moran}, {Moriyama},
  {Moscibrodzka}, {Mul{\ensuremath{\ddot{}}}ler}, {Nagai}, {Nagar}, {Nakamura},
  {Narayan}, {Narayanan}, {Natarajan}, {Neri}, {Ni}, {Noutsos}, {Okino},
  {Olivares}, {Oyama}, {{\"O}zel}, {Palumbo}, {Patel}, {Pen}, {Pesce},
  {Pi{\'e}tu}, {Plambeck}, {PopStefanija}, {Porth}, {Prather},
  {Preciado-L{\'o}pez}, {Psaltis}, {Pu}, {Ramakrishnan}, {Rao}, {Rawlings},
  {Raymond}, {Rezzolla}, {Ripperda}, {Roelofs}, {Rogers}, {Ros}, {Rose},
  {Roshanineshat}, {Rottmann}, {Roy}, {Ruszczyk}, {Ryan}, {Rygl},
  {S{\'a}nchez}, {S{\'a}nchez-Arguelles}, {Sasada}, {Savolainen}, {Schloerb},
  {Schuster}, {Shao}, {Shen}, {Small}, {Sohn}, {SooHoo}, {Tazaki}, {Tiede},
  {Tilanus}, {Titus}, {Toma}, {Torne}, {Trent}, {Trippe}, {Tsuda}, {van
  Bemmel}, {van Langevelde}, {van Rossum}, {Wagner}, {Wardle}, {Weintroub},
  {Wex}, {Wharton}, {Wielgus}, {Wong}, {Wu}, {Young}, {Young}, {Younsi},
  {Yuan}, {Yuan}, {Zensus}, {Zhao}, {Zhao}, {Zhu}, {Anczarski}, {Baganoff},
  {Eckart}, {Farah}, {Haggard}, {Meyer-Zhao}, {Michalik}, {Nadolski},
  {Neilsen}, {Nishioka}, {Nowak}, {Pradel}, {Primiani}, {Souccar},
  {Vertatschitsch}, {Yamaguchi}, \& {Zhang}}]{2019ApJ...875L...5E}
---. 2019{\natexlab{b}}, \apjl, 875, L5

\bibitem[{{Event Horizon Telescope Collaboration} {et~al.}(2021){Event Horizon
  Telescope Collaboration}, {Akiyama}, {Algaba}, {Alberdi}, {Alef}, {Anantua},
  {Asada}, {Azulay}, {Baczko}, {Ball}, {Balokovi{\'c}}, {Barrett}, {Benson},
  {Bintley}, {Blackburn}, {Blundell}, {Boland}, {Bouman}, {Bower}, {Boyce},
  {Bremer}, {Brinkerink}, {Brissenden}, {Britzen}, {Broderick}, {Broguiere},
  {Bronzwaer}, {Byun}, {Carlstrom}, {Chael}, {Chan}, {Chatterjee},
  {Chatterjee}, {Chen}, {Chen}, {Chesler}, {Cho}, {Christian}, {Conway},
  {Cordes}, {Crawford}, {Crew}, {Cruz-Osorio}, {Cui}, {Davelaar}, {De
  Laurentis}, {Deane}, {Dempsey}, {Desvignes}, {Dexter}, {Doeleman}, {Eatough},
  {Falcke}, {Farah}, {Fish}, {Fomalont}, {Ford}, {Fraga-Encinas}, {Freeman},
  {Friberg}, {Fromm}, {Fuentes}, {Galison}, {Gammie}, {Garc{\'\i}a}, {Gentaz},
  {Georgiev}, {Goddi}, {Gold}, {G{\'o}mez}, {G{\'o}mez-Ruiz}, {Gu}, {Gurwell},
  {Hada}, {Haggard}, {Hecht}, {Hesper}, {Ho}, {Ho}, {Honma}, {Huang}, {Huang},
  {Hughes}, {Ikeda}, {Inoue}, {Issaoun}, {James}, {Jannuzi}, {Janssen},
  {Jeter}, {Jiang}, {Jimenez-Rosales}, {Johnson}, {Jorstad}, {Jung}, {Karami},
  {Karuppusamy}, {Kawashima}, {Keating}, {Kettenis}, {Kim}, {Kim}, {Kim},
  {Kim}, {Kino}, {Koay}, {Kofuji}, {Koch}, {Koyama}, {Kramer}, {Kramer},
  {Krichbaum}, {Kuo}, {Lauer}, {Lee}, {Levis}, {Li}, {Li}, {Lindqvist}, {Lico},
  {Lindahl}, {Liu}, {Liu}, {Liuzzo}, {Lo}, {Lobanov}, {Loinard}, {Lonsdale},
  {Lu}, {MacDonald}, {Mao}, {Marchili}, {Markoff}, {Marrone}, {Marscher},
  {Mart{\'\i}-Vidal}, {Matsushita}, {Matthews}, {Medeiros}, {Menten}, {Mizuno},
  {Mizuno}, {Moran}, {Moriyama}, {Moscibrodzka}, {M{\"u}ller}, {Musoke},
  {Mej{\'\i}as}, {Michalik}, {Nadolski}, {Nagai}, {Nagar}, {Nakamura},
  {Narayan}, {Narayanan}, {Natarajan}, {Nathanail}, {Neilsen}, {Neri}, {Ni},
  {Noutsos}, {Nowak}, {Okino}, {Olivares}, {Ortiz-Le{\'o}n}, {Oyama},
  {{\"O}zel}, {Palumbo}, {Park}, {Patel}, {Pen}, {Pesce}, {Pi{\'e}tu},
  {Plambeck}, {PopStefanija}, {Porth}, {P{\"o}tzl}, {Prather},
  {Preciado-L{\'o}pez}, {Psaltis}, {Pu}, {Ramakrishnan}, {Rao}, {Rawlings},
  {Raymond}, {Rezzolla}, {Ricarte}, {Ripperda}, {Roelofs}, {Rogers}, {Ros},
  {Rose}, {Roshanineshat}, {Rottmann}, {Roy}, {Ruszczyk}, {Rygl},
  {S{\'a}nchez}, {S{\'a}nchez-Arguelles}, {Sasada}, {Savolainen}, {Schloerb},
  {Schuster}, {Shao}, {Shen}, {Small}, {Sohn}, {SooHoo}, {Sun}, {Tazaki},
  {Tetarenko}, {Tiede}, {Tilanus}, {Titus}, {Toma}, {Torne}, {Trent},
  {Traianou}, {Trippe}, {van Bemmel}, {van Langevelde}, {van Rossum}, {Wagner},
  {Ward-Thompson}, {Wardle}, {Weintroub}, {Wex}, {Wharton}, {Wielgus}, {Wong},
  {Wu}, {Yoon}, {Young}, {Young}, {Younsi}, {Yuan}, {Yuan}, {Zensus}, {Zhao},
  \& {Zhao}}]{2021ApJ...910L..12E}
{Event Horizon Telescope Collaboration}, {Akiyama}, K., {Algaba}, J.~C.,
  {et~al.} 2021, \apjl, 910, L12

\bibitem[{{Gammie} {et~al.}(2003){Gammie}, {McKinney}, \&
  {T{\'o}th}}]{2003ApJ...589..444G}
{Gammie}, C.~F., {McKinney}, J.~C., \& {T{\'o}th}, G. 2003, \apj, 589, 444

\bibitem[{{Gammie} \& {Popham}(1998)}]{1998ApJ...498..313G}
{Gammie}, C.~F., \& {Popham}, R. 1998, \apj, 498, 313

\bibitem[{{Hakim}(2011)}]{2011irsm.book.....H}
{Hakim}, R.~J. 2011, {Introduction to Relativistic Statistical Mechanics:
  Classical and Quantum}, doi:10.1142/7881

\bibitem[{{Howes}(2010)}]{2010MNRAS.409L.104H}
{Howes}, G.~G. 2010, \mnras, 409, L104

\bibitem[{{Howes}(2011)}]{2011ApJ...738...40H}
---. 2011, \apj, 738, 40

\bibitem[{{Howes} {et~al.}(2006){Howes}, {Cowley}, {Dorland}, {Hammett},
  {Quataert}, \& {Schekochihin}}]{2006ApJ...651..590H}
{Howes}, G.~G., {Cowley}, S.~C., {Dorland}, W., {et~al.} 2006, \apj, 651, 590

\bibitem[{{Kawazura} {et~al.}(2022){Kawazura}, {Schekochihin}, {Barnes},
  {Dorland}, \& {Balbus}}]{2022JPlPh..88c9011K}
{Kawazura}, Y., {Schekochihin}, A.~A., {Barnes}, M., {Dorland}, W., \&
  {Balbus}, S.~A. 2022, Journal of Plasma Physics, 88, 905880311

\bibitem[{{Kawazura} {et~al.}(2020){Kawazura}, {Schekochihin}, {Barnes},
  {TenBarge}, {Tong}, {Klein}, \& {Dorland}}]{2020PhRvX..10d1050K}
{Kawazura}, Y., {Schekochihin}, A.~A., {Barnes}, M., {et~al.} 2020, Physical
  Review X, 10, 041050

\bibitem[{{Kempski} {et~al.}(2019){Kempski}, {Quataert}, {Squire}, \&
  {Kunz}}]{2019MNRAS.486.4013K}
{Kempski}, P., {Quataert}, E., {Squire}, J., \& {Kunz}, M.~W. 2019, \mnras,
  486, 4013

\bibitem[{{Kunz} {et~al.}(2018){Kunz}, {Abel}, {Klein}, \&
  {Schekochihin}}]{2018JPlPh..84b7101K}
{Kunz}, M.~W., {Abel}, I.~G., {Klein}, K.~G., \& {Schekochihin}, A.~A. 2018,
  Journal of Plasma Physics, 84, 715840201

\bibitem[{{Mahadevan} {et~al.}(1996){Mahadevan}, {Narayan}, \&
  {Yi}}]{1996ApJ...465..327M}
{Mahadevan}, R., {Narayan}, R., \& {Yi}, I. 1996, \apj, 465, 327

\bibitem[{{Mo{\'s}cibrodzka} {et~al.}(2016){Mo{\'s}cibrodzka}, {Falcke}, \&
  {Shiokawa}}]{Moscibrodzka2016}
{Mo{\'s}cibrodzka}, M., {Falcke}, H., \& {Shiokawa}, H. 2016, \aap, 586, A38

\bibitem[{{Narayan} \& {Yi}(1994{\natexlab{a}})}]{1994ApJ...428L..13N}
{Narayan}, R., \& {Yi}, I. 1994{\natexlab{a}}, \apjl, 428, L13

\bibitem[{{Narayan} \& {Yi}(1994{\natexlab{b}})}]{Narayan1994}
---. 1994{\natexlab{b}}, \apjl, 428, L13

\bibitem[{{Narayan} \& {Yi}(1995{\natexlab{a}})}]{Narayan1995a}
---. 1995{\natexlab{a}}, \apj, 444, 231

\bibitem[{{Narayan} \& {Yi}(1995{\natexlab{b}})}]{Narayan1995b}
---. 1995{\natexlab{b}}, \apj, 452, 710

\bibitem[{{{\"O}zel} {et~al.}(2022){{\"O}zel}, {Psaltis}, \&
  {Younsi}}]{2022ApJ...941...88}
{{\"O}zel}, F., {Psaltis}, D., \& {Younsi}, Z. 2022, \apj, 941, 88

\bibitem[{{Porth} {et~al.}(2019){Porth}, {Chatterjee}, {Narayan}, {Gammie},
  {Mizuno}, {Anninos}, {Baker}, {Bugli}, {Chan}, {Davelaar}, {Del Zanna},
  {Etienne}, {Fragile}, {Kelly}, {Liska}, {Markoff}, {McKinney}, {Mishra},
  {Noble}, {Olivares}, {Prather}, {Rezzolla}, {Ryan}, {Stone}, {Tomei},
  {White}, {Younsi}, {Akiyama}, {Alberdi}, {Alef}, {Asada}, {Azulay}, {Baczko},
  {Ball}, {Balokovi{\'c}}, {Barrett}, {Bintley}, {Blackburn}, {Boland},
  {Bouman}, {Bower}, {Bremer}, {Brinkerink}, {Brissenden}, {Britzen},
  {Broderick}, {Broguiere}, {Bronzwaer}, {Byun}, {Carlstrom}, {Chael},
  {Chatterjee}, {Chen}, {Chen}, {Cho}, {Christian}, {Conway}, {Cordes},
  {Geoffrey}, {Crew}, {Cui}, {De Laurentis}, {Deane}, {Dempsey}, {Desvignes},
  {Doeleman}, {Eatough}, {Falcke}, {Fish}, {Fomalont}, {Fraga-Encinas},
  {Freeman}, {Friberg}, {Fromm}, {G{\'o}mez}, {Galison}, {Garc{\'\i}a},
  {Gentaz}, {Georgiev}, {Goddi}, {Gold}, {Gu}, {Gurwell}, {Hada}, {Hecht},
  {Hesper}, {Ho}, {Ho}, {Honma}, {Huang}, {Huang}, {Hughes}, {Ikeda}, {Inoue},
  {Issaoun}, {James}, {Jannuzi}, {Janssen}, {Jeter}, {Jiang}, {Johnson},
  {Jorstad}, {Jung}, {Karami}, {Karuppusamy}, {Kawashima}, {Keating},
  {Kettenis}, {Kim}, {Kim}, {Kim}, {Kino}, {Koay}, {Patrick}, {Koch}, {Koyama},
  {Kramer}, {Kramer}, {Krichbaum}, {Kuo}, {Lauer}, {Lee}, {Li}, {Li},
  {Lindqvist}, {Liu}, {Liuzzo}, {Lo}, {Lobanov}, {Loinard}, {Lonsdale}, {Lu},
  {MacDonald}, {Mao}, {Marrone}, {Marscher}, {Mart{\'\i}-Vidal}, {Matsushita},
  {Matthews}, {Medeiros}, {Menten}, {Mizuno}, {Moran}, {Moriyama},
  {Moscibrodzka}, {M{\"u}ller}, {Nagai}, {Nagar}, {Nakamura}, {Narayanan},
  {Natarajan}, {Neri}, {Ni}, {Noutsos}, {Okino}, {Oyama}, {{\"O}zel},
  {Palumbo}, {Patel}, {Pen}, {Pesce}, {Pi{\'e}tu}, {Plambeck}, {PopStefanija},
  {Preciado-L{\'o}pez}, {Psaltis}, {Pu}, {Ramakrishnan}, {Rao}, {Rawlings},
  {Raymond}, {Ripperda}, {Roelofs}, {Rogers}, {Ros}, {Rose}, {Roshanineshat},
  {Rottmann}, {Roy}, {Ruszczyk}, {Rygl}, {S{\'a}nchez},
  {S{\'a}nchez-Arguelles}, {Sasada}, {Savolainen}, {Schloerb}, {Schuster},
  {Shao}, {Shen}, {Small}, {Sohn}, {SooHoo}, {Tazaki}, {Tiede}, {Tilanus},
  {Titus}, {Toma}, {Torne}, {Trent}, {Trippe}, {Tsuda}, {van Bemmel}, {van
  Langevelde}, {van Rossum}, {Wagner}, {Wardle}, {Weintroub}, {Wex}, {Wharton},
  {Wielgus}, {Wong}, {Wu}, {Young}, {Young}, {Yuan}, {Yuan}, {Zensus}, {Zhao},
  {Zhao}, {Zhu}, \& {Event Horizon Telescope Collaboration}}]{Porth2019}
{Porth}, O., {Chatterjee}, K., {Narayan}, R., {et~al.} 2019, \apjs, 243, 26

\bibitem[{{Quataert}(1998)}]{1998ApJ...500..978Q}
{Quataert}, E. 1998, \apj, 500, 978

\bibitem[{{Quataert} \& {Gruzinov}(1999)}]{1999ApJ...520..248Q}
{Quataert}, E., \& {Gruzinov}, A. 1999, \apj, 520, 248

\bibitem[{{Ressler} {et~al.}(2015){Ressler}, {Tchekhovskoy}, {Quataert},
  {Chandra}, \& {Gammie}}]{2015MNRAS.454.1848R}
{Ressler}, S.~M., {Tchekhovskoy}, A., {Quataert}, E., {Chandra}, M., \&
  {Gammie}, C.~F. 2015, \mnras, 454, 1848

\bibitem[{{Ressler} {et~al.}(2017){Ressler}, {Tchekhovskoy}, {Quataert}, \&
  {Gammie}}]{2017MNRAS.467.3604R}
{Ressler}, S.~M., {Tchekhovskoy}, A., {Quataert}, E., \& {Gammie}, C.~F. 2017,
  \mnras, 467, 3604

\bibitem[{{Rowan} {et~al.}(2017){Rowan}, {Sironi}, \&
  {Narayan}}]{2017ApJ...850...29R}
{Rowan}, M.~E., {Sironi}, L., \& {Narayan}, R. 2017, \apj, 850, 29

\bibitem[{{Rowan} {et~al.}(2019){Rowan}, {Sironi}, \&
  {Narayan}}]{2019ApJ...873....2R}
---. 2019, \apj, 873, 2

\bibitem[{{Satapathy} {et~al.}(2022){Satapathy}, {Psaltis}, {{\"O}zel},
  {Medeiros}, {Dougall}, {Chan}, {Wielgus}, {Prather}, {Wong}, {Gammie},
  {Akiyama}, {Alberdi}, {Alef}, {Algaba}, {Anantua}, {Asada}, {Azulay},
  {Baczko}, {Ball}, {Balokovi{\'c}}, {Barrett}, {Benson}, {Bintley},
  {Blackburn}, {Blundell}, {Boland}, {Bouman}, {Bower}, {Boyce}, {Bremer},
  {Brinkerink}, {Brissenden}, {Britzen}, {Broderick}, {Broguiere}, {Bronzwaer},
  {Bustamente}, {Byun}, {Carlstrom}, {Chael}, {Chatterjee}, {Chatterjee},
  {Chen}, {Chen}, {Cho}, {Christian}, {Conway}, {Cordes}, {Crawford}, {Crew},
  {Cruz-Osorio}, {Cui}, {Davelaar}, {De Laurentis}, {Deane}, {Dempsey},
  {Desvignes}, {Dexter}, {Doeleman}, {Eatough}, {Falcke}, {Farah}, {Fish},
  {Fomalont}, {Ford}, {Fraga-Encinas}, {Friberg}, {Fromm}, {Fuentes},
  {Galison}, {Garc{\'\i}a}, {Gentaz}, {Georgiev}, {Goddi}, {Gold},
  {G{\'o}mez-Ruiz}, {G{\'o}mez}, {Gu}, {Gurwell}, {Hada}, {Haggard}, {Hecht},
  {Hesper}, {Ho}, {Ho}, {Honma}, {Huang}, {Huang}, {Hughes}, {Ikeda}, {Inoue},
  {Issaoun}, {James}, {Jannuzi}, {Janssen}, {Jeter}, {Jiang},
  {Jimenez-Rosales}, {Johnson}, {Jorstad}, {Jung}, {Karami}, {Karuppusamy},
  {Kawashima}, {Keating}, {Kettenis}, {Kim}, {Kim}, {Kim}, {Kim}, {Kino},
  {Koay}, {Kofuji}, {Koch}, {Koyama}, {Kramer}, {Kramer}, {Krichbaum}, {Kuo},
  {Lauer}, {Lee}, {Levis}, {Li}, {Li}, {Lindqvist}, {Lico}, {Lindahl}, {Liu},
  {Liu}, {Liuzzo}, {Lo}, {Lobanov}, {Loinard}, {Lonsdale}, {Lu}, {MacDonald},
  {Mao}, {Marchili}, {Markoff}, {Marrone}, {Marscher}, {Mart{\'\i}-Vidal},
  {Matsushita}, {Matthews}, {Menten}, {Mizuno}, {Mizuno}, {Moran}, {Moriyama},
  {Moscibrodzka}, {M{\"u}ller}, {Mej{\'\i}as}, {Musoke}, {Nagai}, {Nagar},
  {Nakamura}, {Narayan}, {Narayanan}, {Natarajan}, {Nathanail}, {Neilsen},
  {Neri}, {Ni}, {Noutsos}, {Nowak}, {Okino}, {Olivares}, {Ortiz-Le{\'o}n},
  {Oyama}, {Palumbo}, {Park}, {Patel}, {Pen}, {Pesce}, {Pi{\'e}tu}, {Plambeck},
  {PopStefanija}, {Porth}, {P{\"o}tzl}, {Preciado-L{\'o}pez}, {Pu},
  {Ramakrishnan}, {Rao}, {Rawlings}, {Raymond}, {Rezzolla}, {Ripperda},
  {Roelofs}, {Rogers}, {Ros}, {Rose}, {Roshanineshat}, {Rottmann}, {Roy},
  {Ruszczyk}, {Rygl}, {S{\'a}nchez}, {S{\'a}nchez-Arguelles}, {Sasada},
  {Savolainen}, {Schloerb}, {Schuster}, {Shao}, {Shen}, {Small}, {Sohn},
  {SooHoo}, {Sun}, {Tazaki}, {Tetarenko}, {Tiede}, {Tilanus}, {Titus}, {Toma},
  {Torne}, {Traianou}, {Trent}, {Trippe}, {van Bemmel}, {van Langevelde}, {van
  Rossum}, {Wagner}, {Ward-Thompson}, {Wardle}, {Weintroub}, {Wex}, {Wharton},
  {Wiik}, {Wu}, {Yoon}, {Young}, {Young}, {Younsi}, {Yuan}, {Yuan}, {Zensus},
  {Zhao}, \& {Zhao}}]{2022ApJ...925...13S}
{Satapathy}, K., {Psaltis}, D., {{\"O}zel}, F., {et~al.} 2022, \apj, 925, 13

\bibitem[{{Schekochihin} {et~al.}(2008){Schekochihin}, {Cowley}, {Dorland},
  {Hammett}, {Howes}, {Plunk}, {Quataert}, \& {Tatsuno}}]{2008PPCF...50l4024S}
{Schekochihin}, A.~A., {Cowley}, S.~C., {Dorland}, W., {et~al.} 2008, Plasma
  Physics and Controlled Fusion, 50, 124024

\bibitem[{{Schekochihin} {et~al.}(2019){Schekochihin}, {Kawazura}, \&
  {Barnes}}]{2019JPlPh..85c9003S}
{Schekochihin}, A.~A., {Kawazura}, Y., \& {Barnes}, M.~A. 2019, Journal of
  Plasma Physics, 85, 905850303

\bibitem[{{Sharma} {et~al.}(2007){Sharma}, {Quataert}, {Hammett}, \&
  {Stone}}]{2007ApJ...667..714S}
{Sharma}, P., {Quataert}, E., {Hammett}, G.~W., \& {Stone}, J.~M. 2007, \apj,
  667, 714

\bibitem[{{Sironi} \& {Narayan}(2015)}]{2015ApJ...800...88S}
{Sironi}, L., \& {Narayan}, R. 2015, \apj, 800, 88

\bibitem[{{S{\k{a}}dowski} {et~al.}(2015){S{\k{a}}dowski}, {Narayan},
  {Tchekhovskoy}, {Abarca}, {Zhu}, \& {McKinney}}]{Sadowski2015}
{S{\k{a}}dowski}, A., {Narayan}, R., {Tchekhovskoy}, A., {et~al.} 2015, \mnras,
  447, 49

\bibitem[{{S{\k{a}}dowski} {et~al.}(2017){S{\k{a}}dowski}, {Wielgus},
  {Narayan}, {Abarca}, {McKinney}, \& {Chael}}]{Sadowski2017}
{S{\k{a}}dowski}, A., {Wielgus}, M., {Narayan}, R., {et~al.} 2017, \mnras, 466,
  705

\bibitem[{{Xie} {et~al.}(2023){Xie}, {Narayan}, \& {Yuan}}]{Xie2023}
{Xie}, F.-G., {Narayan}, R., \& {Yuan}, F. 2023, \apj, 942, 20

\bibitem[{{Yuan} \& {Narayan}(2014)}]{2014ARA&A..52..529Y}
{Yuan}, F., \& {Narayan}, R. 2014, \araa, 52, 529

\end{thebibliography}
\end{document}